\documentclass[pra,twocolumn,a4paper]{revtex4}
\usepackage{graphicx}
\usepackage{amsmath}
\usepackage{url}
\usepackage{bm}
\usepackage{bbm}
\DeclareMathOperator{\tr}{tr}

\renewcommand{\Re}{\operatorname{Re}}
\newcommand{\tsum}{{\textstyle\sum}}

\begin{document}

\title{Quantum-Logic Gate between Two Optical Photons \\ with an Average Efficiency above 40\%}
\author{Thomas Stolz}
\author{Hendrik Hegels}
\author{Maximilian Winter}
\author{Bianca R\"ohr}
\author{Ya-Fen Hsiao}
\author{Lukas Husel}
\author{Gerhard Rempe}
\author{Stephan D\"urr}
\affiliation{Max-Planck-Institut f\"{u}r Quantenoptik, Hans-Kopfermann-Stra{\ss}e 1, 85748 Garching, Germany}

\begin{abstract}
Optical qubits uniquely combine information transfer in optical fibers with a good processing capability and are therefore attractive tools for quantum technologies. A large challenge, however, is to overcome the low efficiency of two-qubit logic gates. The experimentally achieved efficiency in an optical controlled NOT (CNOT) gate reached approximately 11\% in 2003 and has seen no increase since. Here we report on a new platform that was designed to surpass this long-standing record. The new scheme avoids inherently probabilistic protocols and, instead, combines aspects of two established quantum nonlinear systems: atom-cavity systems and Rydberg electromagnetically induced transparency. We demonstrate a CNOT gate between two optical photons with an average efficiency of 41.7(5)\% at a postselected process fidelity of 81(2)\%. Moreover, we extend the scheme to a CNOT gate with multiple target qubits and produce entangled states of presently up to five photons. All these achievements are promising and have the potential to advance optical quantum information processing in which almost all advanced protocols would profit from high-efficiency logic gates.
\end{abstract}

\maketitle

\section{Introduction}

The field of quantum information processing based on optical systems has advanced impressively in the last two decades \cite{Nielsen:00, OBrien:07, Kok:07, Wehner:18}. A recent highlight is the experimental demonstration of the quantum computational advantage in Gaussian boson sampling \cite{Zhong:20}. However, the low efficiency of optical two-qubit gates hampers many other developments in the field. Hitherto, the record for the experimentally achieved efficiency of approximately 11\% was set in 2003 \cite{OBrien:03} with a scheme that uses only linear optics and postselection. Without additional resources, such inherently probabilistic schemes experience a fundamental upper bound for the efficiency of 1/9 \cite{Kieling:10}. With additional resources, including a large number of ancilla photons from deterministic single-photon sources, the KLM scheme \cite{Knill:01} shows that it is hypothetically possible to get arbitrarily close to 100\% efficiency. In practice, however, and despite experimental progress regarding such improved schemes \cite{Li:21}, the actually achieved efficiency of a controlled NOT (CNOT) gate between optical photons has not exceeded 1/9 yet. An alternative strategy focuses on CNOT gates based on quantum nonlinear systems that avoid inherently probabilistic protocols. Such gates were demonstrated in two recent experiments, one using an atom-cavity system \cite{Hacker:16}, the other using Rydberg electromagnetically induced transparency (EIT) \cite{Tiarks:19}. However, the demonstrated average efficiency in these two experiments remained below 5\% for technical reasons.

We combine key components of these two experiments in a cavity Rydberg EIT experiment \cite{Hao:15, Das:16, Pritchard:10, Parigi:12, Firstenberg:13, Jia:18}. This new approach overcomes previous difficulties and, for the first time, outperforms the 1/9 bound of postselected linear optical systems. It improves the long-standing record for the average efficiency of optical CNOT gates by a factor of 3.8, leaving a factor of 2.4 to 100\% efficiency. An analysis of current imperfections shows that these can be overcome in the future. Moreover, theory predicts that much higher efficiencies and fidelities are realistic with our system. This opens up perspectives for applications in optical quantum computing \cite{OBrien:07}, in distributed quantum computing \cite{Gyongyosi:21}, or in high-efficiency optical Bell-state detection \cite{Calsamiglia:01}, which would be useful for linear-optics quantum-repeater schemes \cite{Guha:15} and for a future quantum internet \cite{Wehner:18}. Concepts might also be transferred to the microwave domain \cite{Reuer:22}. Our multiple-target CNOT gate is interesting for applications in quantum error correction \cite{Nielsen:00}, and multiphoton entangled states might become powerful resources for linear optical quantum computing \cite{Kok:07} and for measurement-based quantum computing \cite{Raussendorf:01}.

\section{Protocol}

\begin{figure}[!tb]
\centering
\includegraphics[width=\columnwidth]{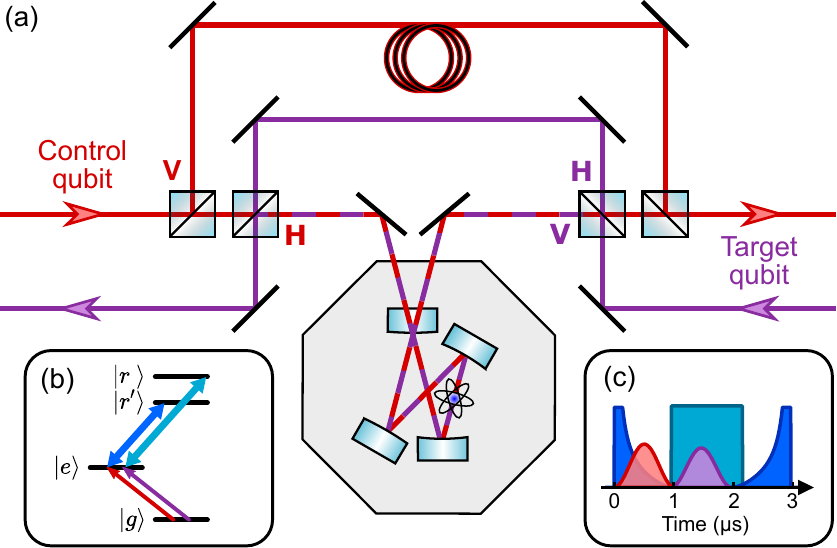}
\caption{(a) Scheme of the gate. The control (target) photon pulse travels through the setup from left to right (right to left). Polarizing beam splitters (PBSs, blue squares) map between incoming and outgoing polarization qubits and internal dual-rail qubits. One rail of each qubit impinges on the cavity, and the other rail bypasses the cavity. If the control qubit impinges on the cavity, it is stored in a Rydberg state. Subsequently, the target photon enters the system. A conditional $\pi$ phase shift is experienced if both qubits are in their cavity rails. After the interaction with the target photon, the control photon is retrieved. The bypass rail of the control qubit is delayed in an optical fiber to match the delay resulting from storage. The octagon represents the vacuum chamber. (b) Atomic level scheme. Coupling light (blue, cyan) creates EIT for the signal light (red, purple). (c) Timing sequence of the light power of the incoming pulses. The vertical axis uses different scales for different light fields. In the first microsecond, the control pulse is stored; in the second microsecond, the target pulse interacts with the system; and in the third microsecond, the control pulse is retrieved.}
\label{fig-scheme}
\end{figure}

Rydberg EIT in an ultracold atomic gas essentially maps the giant dipole-dipole interaction between Rydberg atoms onto optical photons. This makes it possible to build a single-photon switch \cite{Baur:14, Tiarks:14, Gorniaczyk:14, Gorniaczyk:16} in which the presence of a single photon toggles the transmission of another photon from on to off. To this end, it is advantageous to store the first photon, called the control photon, in the atomic ensemble in the atomic Rydberg state $|r'\rangle$, see Fig.\ \ref{fig-scheme}(b), using EIT-based storage \cite{Gorshkov:07:cavity}. Next, the second photon, called the target photon, impinges on the atomic ensemble. Finally, the control photon is retrieved, see Fig.\ \ref{fig-scheme}(c) for an overview of this timing sequence.

In the absence of a stored excitation in state $|r'\rangle$, the target photon experiences Rydberg EIT with a Rydberg state $|r\rangle$ resulting in high transmission. If an excitation in state $|r'\rangle$ is present, however, then the interaction between the two Rydberg states will shift the system away from the EIT two-photon resonance, thus causing low transmission of the target photon. Overall, this yields loss of the target photon conditioned on the presence of a single control photon.

Placing this system inside a one-sided optical resonator converts the conditional loss into a conditional $\pi$ phase shift \cite{Hofmann:03, Duan:04, Hacker:16}. This is because, for large (small) intra-cavity loss, the cavity is undercoupled (overcoupled). This cavity approach offers an efficient way of harvesting the nonlinearity of Rydberg EIT and makes a high-efficiency CNOT gate possible, similar to the proposals in Refs.\ \cite{Hao:15, Das:16}. Compared to our previous work \cite{Tiarks:19}, this cavity approach offers the additional advantage that the conditional $\pi$ phase shift is much more robust against changes in the optical depth of the atomic ensemble.

\section{Experimental Setup}

Figure \ref{fig-scheme}(a) shows how we convert this conditional $\pi$ phase shift into a controlled $\pi$ phase (CPHASE) gate for the polarization qubits of two optical photons. Ideally, the gate should be a linear map of state vectors onto state vectors, characterized by
\begin{subequations}
\begin{align}
\label{ideal-gate-a}%
|HH\rangle
&
\mapsto |HH\rangle
, &
|HV\rangle
&
\mapsto e^{i\pi}|HV\rangle
, \\
\label{ideal-gate-b}%
|VH\rangle
&
\mapsto |VH\rangle
, &
|VV\rangle
&
\mapsto |VV\rangle
.\end{align}
\end{subequations}
Here, the control qubit is listed first and $H$, $V$, $D$, and $A$ denote the linear polarizations which are horizontal ($H$), vertical ($V$), diagonal ($D, 45^\circ$), and antidiagonal ($A, -45^\circ$). In essence, the scheme in the figure temporarily converts the incoming polarization qubits into dual-rail qubits. The quantum amplitude corresponding to the rails in which both qubits impinge on the cavity experiences the conditional $\pi$ phase shift. The CPHASE gate is identical to a CNOT gate up to single-qubit unitaries \cite{Nielsen:00}, which are easily implemented using wave plates.

In our experiment, the incoming signal light pulses are derived from a laser and have a Poissonian photon number distribution with a mean photon number much below 1. Some data in this paper are postselected upon the detection of exactly one control photon and exactly one target photon. Mostly, this removes cases in which an incoming pulse contains zero photons or in which an incoming pulse contains one photon which is lost. Note, however, that the gate does not use an inherently probabilistic scheme.

The core of the experiment is an ensemble of 260 $^{87}$Rb atoms at a temperature of 0.4 $\mu$K, held in an optical dipole trap (ODT), located in a bow-tie cavity with a finesse of 350. The coupled atom-cavity system is characterized by $(g,\kappa,\gamma)/2\pi= (1.0,2.3,3.0)$ MHz, where $g$ is half of the vacuum Rabi frequency for a maximally-coupled single atom, $\kappa$ is the half width at half maximum (HWHM) cavity linewidth, and $\gamma$ is the HWHM linewidth of the atomic transition $|g\rangle\leftrightarrow |e\rangle$. This results in a collective cooperativity of 21. Each light pulse lasts 1 $\mu$s, thus roughly matching $2\pi/\kappa$. The experiment is repeated every 100 $\mu$s. For details regarding the setup, see appendix \ref{app-setup}.

\section{Results}

\subsection{Initial Characterization}

\begin{figure}[!tb]
\centering
\includegraphics[width=\columnwidth]{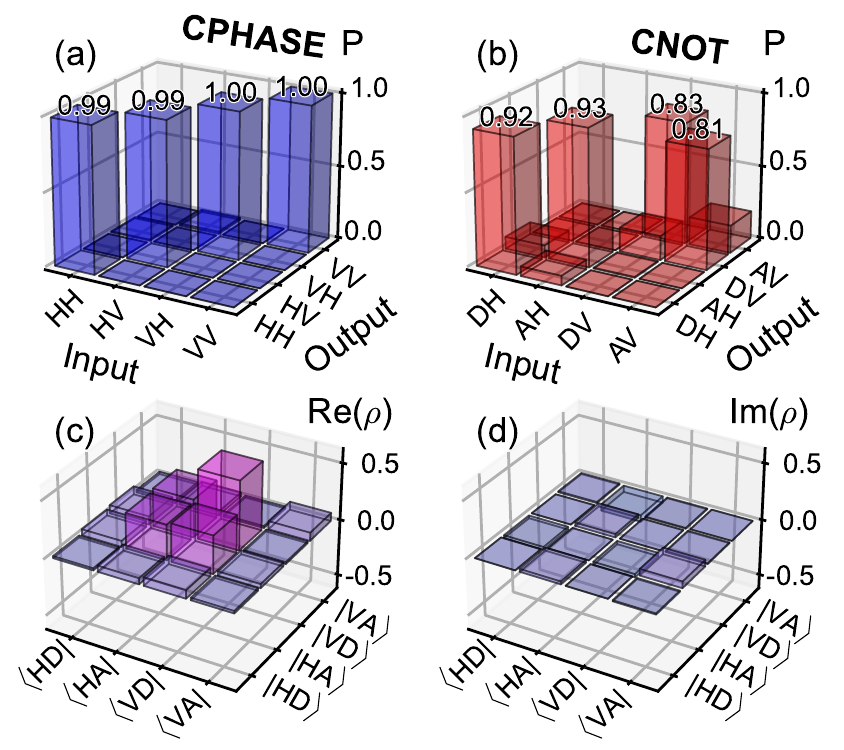}
\caption{Truth tables and Bell-state tomography. (a) Postselected truth table in the CPHASE basis. The average fidelity is 99.4(4)\%. All error bars in this paper represent a statistical uncertainty of 1 standard deviation. (b) Postselected truth table in a CNOT basis. The average fidelity is 87(1)\%. (c,d) Real and imaginary parts of the reconstructed postselected density matrix obtained in an entangling gate operation. The postselected Bell-state fidelity is 79(2)\%.}
\label{fig-truth-tables}
\end{figure}

A widely used method \cite{OBrien:03, Hacker:16, Tiarks:19} for a first characterization of a CNOT gate is to consider a CNOT truth table together with one generated Bell state. Additionally, we consider a CPHASE truth table. Figures \ref{fig-truth-tables}(a) and \ref{fig-truth-tables}(b) show postselected truth tables in the CPHASE basis and a CNOT basis. To measure these data, one of the indicated states is used as an input, and the measured probabilities of obtaining the indicated output states in a measurement are displayed. The average fidelity of the truth table is the arithmetic mean of the four labeled probabilities.

The postselected CPHASE truth table in Fig.\ \ref{fig-truth-tables}(a) is interesting because it is very close to ideal, thus excluding a variety of otherwise possible sources of experimental imperfections. This insight will be useful below. This is expected from the setup because photons do not jump from one rail to the other. Hence, deviations from a perfect CPHASE truth table come mostly from imperfections in PBSs and wave plates.

Figures \ref{fig-truth-tables}(c) and \ref{fig-truth-tables}(d) show the postselected output density matrix obtained in quantum-state tomography \cite{Nielsen:00} for the input state $|DD\rangle$. Ideally, the output state should be a Bell state. The fact that the postselected Bell-state fidelity \cite{Gilchrist:05} exceeds 50\% amounts to witnessing two-qubit entanglement. The postselected Bell-state fidelity obtained here is clearly higher than the 64\% in Ref.\ \cite{Tiarks:19} and a bit higher than the 76\% in Ref.\ \cite{Hacker:16}.

\subsection{Quantum Process Tomography}

\label{sec-tomography}

Instead of characterizing the output state for only a few selected input states, we now fully analyze the performance using quantum process tomography \cite{Nielsen:00}. The latter gives a complete characterization of the quantum process, i.e.\ it yields a map $\mathcal E$ from the $4\times 4$ density matrix $\rho_\text{in}$ at the input to the $4\times 4$ density matrix $\rho_\text{out} = \mathcal E(\rho_\text{in})$ at the output. The map $\mathcal E$ is assumed to be linear. Hence, to characterize $\mathcal E$, it suffices to experimentally determine $\rho_\text{out}$ using quantum-state tomography for a basis of 16 matrices $\rho_\text{in}$.

Thus, the reconstructed map $\mathcal E$ can be written in a chi-matrix representation as \cite{Nielsen:00}
\begin{align}
\label{E-rho-chi}
\mathcal E(\rho)
= \sum_{i,j=1}^{16} A_i \rho A_j^\dag \chi_{i,j}
,\end{align}
where the $4\times 4$ matrices $A_i$ are chosen at will to form a basis $\mathcal A$ of operators. The complex expansion coefficients $\chi_{i,j}$ form the $16\times 16$ process matrix $\chi$ \cite{Gilchrist:05}. Of course, the $\chi_{i,j}$ depend on the choice of the basis $\mathcal A$. From the experimental data for a basis of 16 matrices $\rho_\text{in}$, we calculate a linear, unbiased estimator for $\chi$ by inverting a linear system \cite{Nielsen:00}, see also appendix \ref{app-tomography}. Once $\chi$ has been determined, $\rho_\text{out}= \mathcal E(\rho_\text{in})$ can be predicted for an arbitrary $\rho_\text{in}$ using Eq.\ \eqref{E-rho-chi}.

Possible loss of photons is represented by $\tr(\rho_\text{out})\leq 1$, i.e.\ $\mathcal E$ is typically not trace preserving \cite{Nielsen:00}. If the basis $\mathcal A$ was orthonormal, i.e.\ if $\tr(A_i^\dag A_j)= \delta_{i,j}$, then one would obtain $\tr(\chi)= 4\bar\eta$ with the average efficiency $\bar\eta$ of Eq.\ \eqref{eta-bar-sum}. We normalize the process matrix upon postselection to correct its trace, i.e.\ we consider the postselected process matrix  $\chi^\text{ps}= \chi/\bar\eta$.

We choose $A_i= U B_i$, where the unitary $4\times 4$ matrix $U$ describes the operation of the ideal two-qubit gate given by Eqs.\ \eqref{ideal-gate-a} and \eqref{ideal-gate-b} and where the $4\times 4$ matrices $B_i$ are the 16 tensor products which can be formed from the single-qubit Pauli matrices $I$, $X$, $Y$, and $Z$ in the $(|H\rangle,|V\rangle)$ basis, i.e.\ $B_1= I\otimes I$, $B_2= I\otimes X$, $B_3= I\otimes Y$, ..., $B_{16}= Z\otimes Z$. The $\frac12A_i$ form an orthonormal basis. Hence $\tr(\chi^\text{ps})= 1$. This choice of the $A_i$ is beneficial because for the ideal process matrix, it yields the simple expression $\chi^\text{id}_{i,j}= \delta_{1,i}\delta_{1,j}$.

\begin{figure}[!tb]
\centering
\includegraphics[width=\columnwidth]{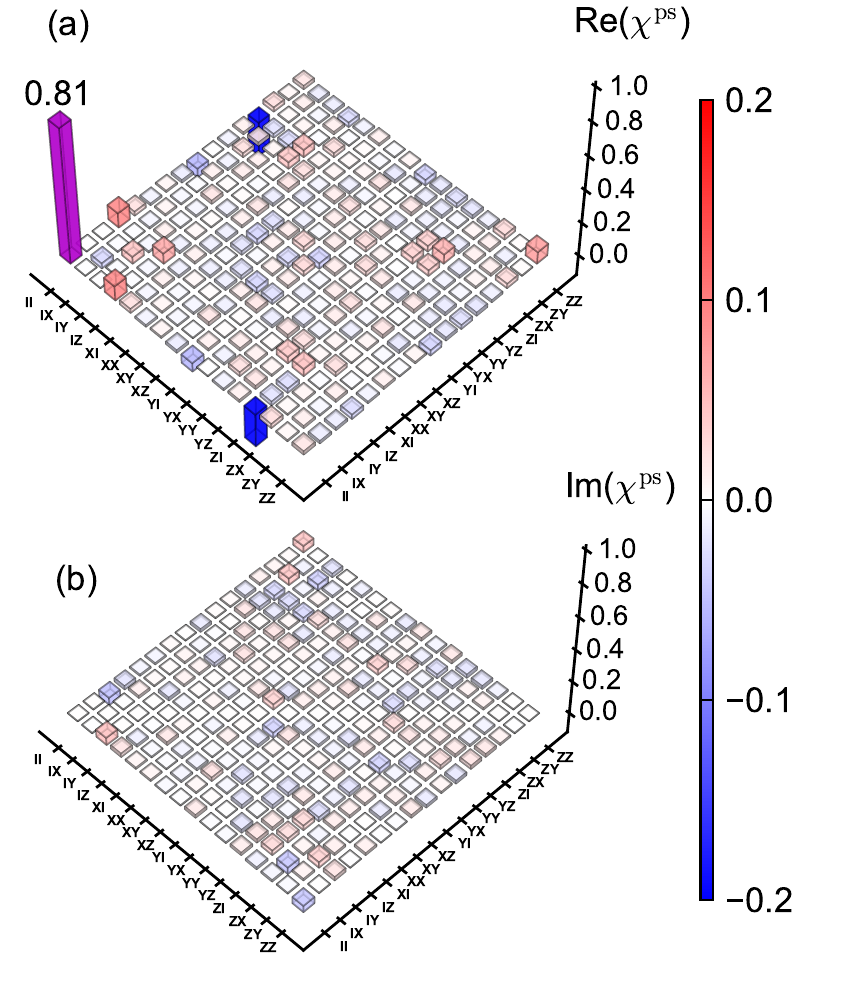}
\caption{Quantum process tomography. Real and imaginary parts of the elements of the postselected $16\times16$ process matrix $\chi^\text{ps}$. The basis for the matrix representation is chosen such that for the ideal gate, one matrix element would equal unity and all others would vanish. The postselected process fidelity is 81(2)\%.}
\label{fig-process-tomography}
\end{figure}

The elements of $\chi^\text{ps}$ are shown in Fig.\ \ref{fig-process-tomography} with the $B_i$ used as labels. A simple model, which includes only the four measured efficiencies of the CPHASE basis states and two single-qubit visibilities $V_c= 86(4)\%$ and $V_t= 78(3)\%$ for the control and the target qubit agrees fairly well with the measured $\chi^\text{ps}$, see appendix \ref{app-process-matrix}.

Now, we turn to the process fidelity. If the basis $\mathcal A$ was orthonormal, then the process fidelity would be \cite{Gilchrist:05} $F_\text{pro}= \frac1{16} \tr(\chi^\text{id}\chi)$. Hence, $F_\text{pro}= \tr(\chi^\text{id}\chi)$ for the basis $\mathcal A$ chosen here. The postselected version thereof is the postselected process fidelity $F^\text{ps}_\text{pro}= \tr(\chi^\text{id}\chi^\text{ps})$. Using that $\mathcal E$ is typically a completely positive map \cite{Nielsen:00}, one can show that this yields $F^\text{ps}_\text{pro}\leq 1$. Using $\chi^\text{id}_{i,j}= \delta_{1,i}\delta_{1,j}$, one obtains $F^\text{ps}_\text{pro}= \chi^\text{ps}_{1,1}$ \cite{OBrien:04, Riebe:06}. Hence, Fig.\ \ref{fig-process-tomography} yields $F^\text{ps}_\text{pro}=81(2)$\%. This is clearly above the classical limit of 50\%. Multiplying $F^\text{ps}_\text{pro}$ with the average efficiency $\bar\eta$ of Eq.\ \eqref{eta-bar-sum} yields a process fidelity without postselection of $F_\text{pro}= 34(1)\%$, which is much better than in all previous measurements of CNOT gates with optical photons, in which $F_\text{pro}$ never exceeded 11\%.

The postselected Bell-state fidelity in Figs.\ \ref{fig-truth-tables}(c) and (d) equals the postselected process fidelity $F^\text{ps}_\text{pro}$ in Fig.\ \ref{fig-process-tomography} within the experimental uncertainties. This is expected because the postselected CPHASE truth table in Fig.\ \ref{fig-truth-tables}(a) is almost ideal, see appendix \ref{app-process-matrix}.

\subsection{Efficiency}

\label{sec-efficiency}

The efficiency of the gate is the probability that no photon is lost inside the gate, if one control and one target photon impinge on the gate. The gate is the part of the setup shown in Fig.\ \ref{fig-scheme}(a). The efficiency also includes optical elements like lenses and wave plates which are not drawn in Fig.\ \ref{fig-scheme}(a) for simplicity. The only components which are not included in the efficiency are the light source and the detection setup. This makes sense because neither light sources nor detectors need to be cascaded in a sequence of gates.

The efficiency $\eta$ depends on the input density matrix $\rho_\text{in}$ and can be written as $\eta(\rho_\text{in})= \tr[\mathcal E(\rho_\text{in})]$. For comparing different experiments, one can introduce various figures of merit, which attempt to express the most relevant information about the function $\eta$ in just one number. Obvious choices include the maximum, the minimum, and various averages. We define the average efficiency $\bar\eta= \int d\psi \eta(|\psi\rangle\langle\psi|)$, where the integral is over the uniform (Haar) measure \cite{Tilma:02:SUN} of all normalized input state vectors, similar to the definition of the average fidelity \cite{Gilchrist:05}. For any orthonormal basis of state vectors $|v_i\rangle$, the average efficiency can be rewritten as, see appendix \ref{app-efficiency}
\begin{align}
\label{eta-bar-sum}
\bar\eta
= \frac14 \sum_{i=1}^4 \eta(|v_i\rangle\langle v_i|)
.\end{align}

Instead of considering only $\bar\eta$, one can characterize the full function $\eta$ for all $\rho_\text{in}$. As the map from $\rho_\text{in}$ to $\eta(\rho_\text{in})$ is a linear form, it can be written as $\eta(\rho_\text{in})= \tr(\Theta^\dag \rho_\text{in})$ with a $4\times 4$ matrix $\Theta$, which we call the efficiency matrix. As $\eta(\rho_\text{in})$ is real for all Hermitian $\rho_\text{in}$, $\Theta$ is Hermitian. For a full characterization of the efficiency function it suffices to measure $\eta(\rho_\text{in})$ for a basis of 16 input density matrices $\rho_\text{in}$. From this, $\Theta$ can be reconstructed by inverting a linear system, see appendix \ref{app-efficiency}. If a process matrix has already been determined, one finds $\Theta= \sum_{i,j=1}^{16} A_j^\dag A_i \chi_{i,j}$.

In our experiment, the postselected CPHASE truth table in Fig.\ \ref{fig-truth-tables}(a) is ideal, to a good approximation. Hence, the matrix representation of $\Theta$ in the CPHASE basis is diagonal, to a good approximation, see appendix \ref{app-process-matrix}. To give the full information about $\Theta$, it therefore suffices to quote only these diagonal elements, i.e.\ the efficiencies obtained if one of the CPHASE basis states is used as an input state. These are measured to be $(\eta_{HH}, \linebreak[1] \eta_{HV}, \linebreak[1] \eta_{VH}, \linebreak[1] \eta_{VV})=(35.1(7), \linebreak[1] 15.7(1.4), \linebreak[1] 61.1(1), \linebreak[1] 54.9(1.0))\%$. According to Eq.\ \eqref{eta-bar-sum}, the arithmetic mean of these four efficiencies is the average efficiency $\bar\eta= 41.7(5)\%$.

These values are obtained when choosing the experimental parameters to maximize $\bar\eta$. If, instead, one was interested e.g.\ in maximizing the minimum eigenvalue $\eta_\text{min}$ of $\Theta$, one would have to choose different parameters. We find e.g.\ $\eta_\text{min}= 24.4(2.2)\%$ for a target coupling Rabi frequency of 20 MHz and an atom number of 500, resulting in a collective cooperativity of 40. At these parameters, we obtain $(\eta_{HH}, \linebreak[1] \eta_{HV}, \linebreak[1] \eta_{VH}, \linebreak[1] \eta_{VV})=(38.2(7), \linebreak[1] 24.4(2.2), \linebreak[1] 61.1(1), \linebreak[1] 34.4(5))\%$ and $\bar\eta=39.5(6)\%$ along with a Bell-state fidelity of 73(1)\% for the output state generated from the input state $|DD\rangle$.

Experimentally, two-photon coincidences are detected at a rate of 560 s$^{-1}$ per incoming photon pair, see appendix \ref{app-coincidence}. This number is factors of 520 and $1.3\times 10^4$ larger than in Refs.\ \cite{Hacker:16} and \cite{Tiarks:19}, respectively, thus making much more demanding experiments possible in a realistic data acquisition time.

A key factor as to why the average efficiency outperforms that reported in Ref.\ \cite{Hacker:16} is that more atoms create more conditional loss. This allows us to operate at higher collective cooperativity and at much lower cavity finesse, which both mitigate a variety of technical problems such as mirror absorption and mirror scattering relative to mirror transmission. Compared to Ref.\ \cite{Tiarks:19}, the outperformance comes from the cavity that makes the light pass through the same Rydberg blockade volume many times. This enables us to reduce the atomic density and, in consequence, the dephasing rate due to atomic collisions \cite{Baur:14, Schmidt-Eberle:20}.

The average efficiency we achieved is below 100\% for technical reasons. The dominant limitation is dephasing, although of a different origin than in Ref.\ \cite{Tiarks:19}. When EIT coupling light is on (off), dephasing is now dominated by laser phase noise (differential light shifts in the dipole trap). Reducing laser phase noise and operating the dipole trap at a different wavelength \cite{Schmidt-Eberle:20} is expected to massively reduce dephasing. In addition, the delay fiber could be replaced by a high-efficiency quantum memory. A detailed analysis based on the models of Refs.\ \cite{Das:16, Gorshkov:07:cavity} predicts an average efficiency of the two-qubit gate far above 90\% along with a much improved postselected fidelity.

\subsection{Multiphoton Entanglement}

The two-photon gate implemented here is easily extended to a multiple-target CNOT gate \cite{Nielsen:00}, thus allowing us to directly produce multiphoton entanglement. This scalability provided by our scheme is advantageous because fewer resources are required compared to cases where several two-qubit gates are cascaded. In particular, an $N$-photon Greenberger-Horne-Zeilinger (GHZ) state $|\psi_N\rangle= (|H^{\otimes N}\rangle+|V^{\otimes N}\rangle)/\sqrt 2$ can be generated by sending one control photon and $N-1$ target photons onto the gate in an input state with polarization $|D^{\otimes N}\rangle$. We choose to assemble all target photons in one pulse. The target pulse lasts long enough that interactions among target photons are negligible. Hence, each target photon simply acquires a $\pi$ phase shift conditioned on the presence of the same control photon. Hence, this input state is mapped onto the output state $(|H\rangle|A^{\otimes (N-1)}\rangle+|V\rangle|D^{\otimes (N-1)}\rangle)/\sqrt 2$. A simple single-qubit unitary applied to all outgoing target photons converts this into the above GHZ state $|\psi_N\rangle$.

\begin{figure}[!tb]
\centering
\includegraphics[width=\columnwidth]{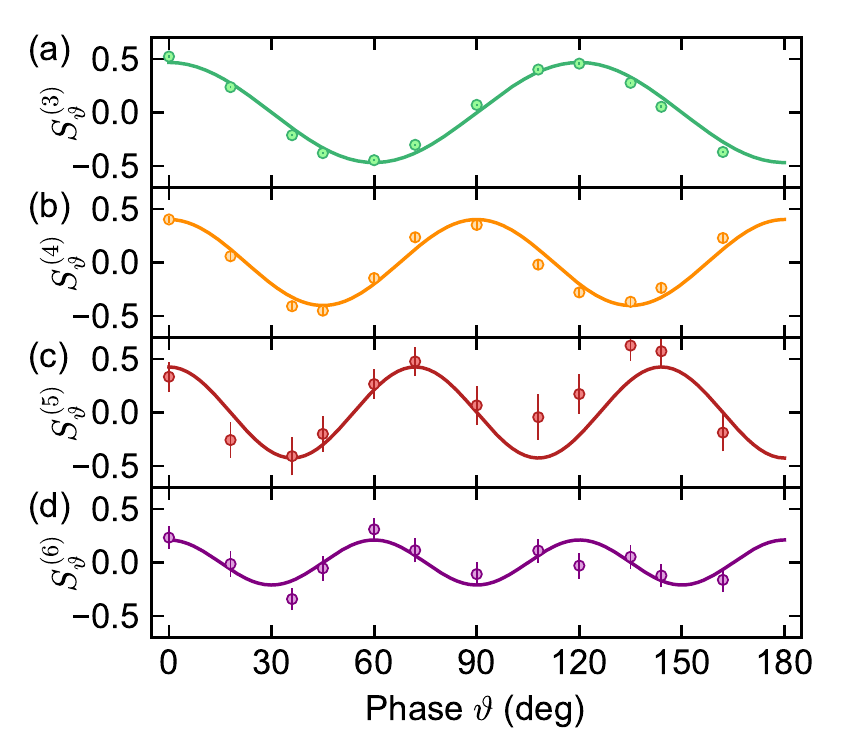}
\caption{Multiphoton entanglement. To verify genuine $N$-photon entanglement, we study parity oscillations. To this end, we measure all photons in the same polarization basis $(|b_{\vartheta,+}\rangle,|b_{\vartheta,-}\rangle)$ with $|b_{\vartheta,\pm}\rangle= (|H\rangle\pm e^{i\vartheta}|V\rangle)/\sqrt 2$, where $\vartheta$ is real. The measurement outcomes yield the generalized Stokes parameter $S_\vartheta^{(N)}= \langle M_\vartheta^{\otimes N}\rangle$, where $M_\vartheta= |b_{\vartheta,+}\rangle\langle b_{\vartheta,+}| - |b_{\vartheta,-}\rangle\langle b_{\vartheta,-}|$ is the single-qubit operator describing the projection of the Stokes vector along a suitable direction. The dots show measured values of $S_\vartheta^{(N)}$ as a function of $\vartheta$ for (a) $N=3$, (b) $N=4$, (c) $N=5$, and (d) $N=6$. The lines are fits, see appendix \ref{app-GHZ}.}
\label{fig-GHZ}
\end{figure}

To detect an $N$-photon GHZ state, we again use input pulses with a Poissonian photon number distribution. But this time, we postselect upon the detection of exactly one control and exactly $N-1$ target photons. To verify the $N$-photon entanglement, we study parity oscillations \cite{Sackett:00, Monz:11, Wang:16}. From the experimental data in Fig.\ \ref{fig-GHZ}, we calculate the coherence $\mathcal C_N = \langle |H^{\otimes N}\rangle\langle V^{\otimes N}| \rangle+\text{c.c.}= \frac1N \sum_{k=0}^{N-1} (-1)^k S_{k\pi/N}^{(N)}$. In addition, we measure the populations $p_H= \langle |H^{\otimes N}\rangle\langle H^{\otimes N}| \rangle$ and $p_V= \langle |V^{\otimes N}\rangle\langle V^{\otimes N}|\rangle$ of the states $|H^{\otimes N}\rangle$ and $|V^{\otimes N}\rangle$ and use the results to calculate $\mathcal P_N= p_H+p_V$. Combining these results, we determine the postselected GHZ-state fidelity \cite{Sackett:00, Monz:11, Wang:16} $F_N= \langle\psi_N|\rho_\text{out}|\psi_N\rangle= \frac12(\mathcal P_N+\mathcal C_N)$ and obtain $F_3= 62.3(4)$\%, $F_4= 54.6(1.4)$\%, $F_5= 54.8(5.3)$\%, and $F_6= 35.9(3.7)$\%. Note that $F_N>50\%$ implies genuine $N$-photon entanglement \cite{Sackett:00, Monz:11, Wang:16}. Assuming that the statistical uncertainties quoted here correspond to a Gaussian distribution, the $p$-value, i.e.\ the probability of $F_N<50\%$, is $(10^{-207}$, $5\times 10^{-4}$, 0.18) for $N=(3, 4, 5)$. A simple model, taking efficiencies and single-qubit visibilities into account agrees well with the experimental data, see appendix \ref{app-GHZ}.

\section{Conclusions}

High-efficiency optical gates have the potential to become useful for many applications in optical quantum information. For example, high-efficiency gates could drastically reduce the resource cost \cite{Li:15} in a variety of optical quantum computing schemes. The fact that the first realization of an optical photon-photon CNOT gate \cite{OBrien:03} already achieved an efficiency around 1/9 and that no improvement had been made in almost two decades might have narrowed the application perspective of optical quantum computing. Questions like which optical quantum computing schemes are most promising might find new answers, now that the present work makes much higher efficiencies in photon-photon quantum gates a reality.

\section*{ACKNOWLEDGEMENTS}

We thank Valentin Walther, Callum Murray, Thomas Pohl, Nathan Schine, Jonathan Simon, Thomas Dieterle, Florian Meinert, Robert L\"ow, Tilman Pfau, Andreas Reiserer, and Stephan Welte for discussions. This work was supported by Deutsche Forschungsgemeinschaft under priority program 1929 GiRyd and under Germany's excellence strategy via Munich Center for Quantum Science and Technology EXC-2111-390814868. T.S.\ acknowledges support from Studienstiftung des deutschen Volkes.

\bigskip
\noindent
\textit{Note added.}---Recently, we became aware of a measurement of a conditional $\pi$ phase shift based on cavity Rydberg EIT \cite{Vaneecloo:22}.

\appendix

\section{Experimental Setup}

\label{app-setup}

\subsection{Cavity}

As the vacuum system used in our previous experiments \cite{Baur:14, Tiarks:14, Tiarks:19} did not offer enough room to house the cavity, we set up a new vacuum system. The cavity consists of four mirrors in bow-tie geometry, see appendix \ref{app-geometry}, with an axial mode spacing of $\Delta\omega_\text{ax}/2\pi= 1.59$ GHz, a finesse of $\mathcal F= 350$, and a beam waist ($1/e^2$ radius of intensity) of $w_c= 8.5$ $\mu$m. The cavity is one-sided, i.e.\ it has an input-output coupler with a moderate reflectivity of $98.3\%$ while each of the other cavity mirrors has a much higher reflectivity of 99.990\%. These numbers refer to EIT signal light at a wavelength of 780 nm. For EIT coupling light at a wavelength of 480 nm, however, all mirrors have negligible reflectivity. A piezo actuator is used to stabilize the cavity length.

\subsection{Atomic Ensemble and Transitions}

The experiment begins with the preparation of an ultracold gas of $^{87}$Rb atoms in a crossed-beam ODT, see appendix \ref{app-cooling}. The atomic ensemble is located at the beam waist of the cavity. We choose the $z$ axis such that it points along the wave vector of the control signal light at the location of the atomic ensemble. We choose the $y$ axis to be vertical. The atomic ensemble has a temperature of $T=0.44$ $\mu$K and an atom number of $N_a= 260$. With measured trapping frequencies of $(\omega_x,\omega_y,\omega_z)/2\pi= (0.31,0.23,0.60)$ kHz, the root-mean-square (rms) radii of the atomic ensemble are estimated to be $(\sigma_x,\sigma_y,\sigma_z)= (3.3,4.5,1.7)$ $\mu$m.

The atomic energy levels used in the experiment are the ground state $|g\rangle= |5S_{1/2},F{=}2,m_F{=}{-}2\rangle$, the first electronically excited state $|e\rangle= |5P_{3/2}, \linebreak[1] F{=}3, \linebreak[1] m_F{=}{-}3\rangle$ and two Rydberg states, namely $|r'\rangle= |48S_{1/2}, \linebreak[1] F{=}2, \linebreak[1] m_F{=}{-}2\rangle$ for the control qubit and $|r\rangle= |50S_{1/2}, \linebreak[1] F{=}2, \linebreak[1] m_F{=}{-}2\rangle$ for the target qubit. Here, $F,m_F$ are hyperfine quantum numbers. All atoms are prepared in the stretched spin state $|g\rangle$. A magnetic hold field of 0.10 mT applied along the positive $z$ axis suppresses the randomization of the spin orientation by noise of the ambient magnetic field.

The rms cloud radii along all three dimensions are smaller than the Rydberg blockade radius of approximately 7 $\mu$m, see appendix \ref{app-blockade}. This results in a superatom geometry \cite{Saffman:10}, in which a single stored Rydberg excitation from the control pulse causes Rydberg blockade for all atoms during the target pulse. To address an electrically tuned F\"orster resonance \cite{Gorniaczyk:16}, an electric field of 0.7 V/cm is applied along the positive $z$ axis.

\subsection{EIT}

The EIT signal light is resonant with the $|g\rangle \leftrightarrow |e\rangle$ transition. The EIT coupling light during the control (target) pulse is resonant with the $|e\rangle \leftrightarrow |r'\rangle$ ($|e\rangle \leftrightarrow |r\rangle$) transition. As seen in Fig.\ \ref{fig-scheme}(a), the EIT signal light beams for control and target light counterpropagate each other. The wave vector of each of the two EIT coupling beams is chosen to counterpropagate the corresponding EIT signal beam. This minimizes the net photon recoil in the EIT two-photon transition. Hence, this minimizes dephasing caused by thermal atomic motion \cite{Schmidt-Eberle:20}. The control (target) EIT coupling beam has a peak power of 30 mW (70 mW). At this value, the coupling Rabi frequency was measured to be 20 MHz (43 MHz), see appendix \ref{app-EIT}. From this, we estimate a beam waist of 21 $\mu$m (14 $\mu$m).

The incoming control signal light pulse has an intensity proportional to $\cos^2(\pi(t-t_{c,\text{in}})/t_c)$ if $|t-t_{c,\text{in}}|\leq t_c/2$ and 0 otherwise. Here, $t_c= 1.0$ $\mu$s is the control pulse duration and $t_{c,\text{in}}$ is the center of the control pulse. The outgoing control signal light pulse has the same pulse shape with the same duration $t_c$ but a delayed center $t_{c,\text{out}}= t_{c,\text{in}}+ t_c+t_t$. The incoming target signal light also has the same pulse shape with a target pulse duration $t_t= 1.0$ $\mu$s and a center of the incoming target pulse $t_{t,\text{in}}= t_{c,\text{in}}+(t_c+t_t)/2$. During storage and retrieval, the pulse shape of the control EIT coupling light is chosen according to the model of Ref.\ \cite{Gorshkov:07:cavity}. During the target pulse, however, the target EIT coupling light has a constant power, which is chosen to be the above-mentioned maximum.

\subsection{Atom-Cavity Coupling}

An important figure of merit characterizing the coupled atom-cavity system is the collective cooperativity $C= \sum_{i=1}^{N_a} |g_i|^2/\kappa\gamma$, where $g_i$ is half of the vacuum Rabi frequency of the $i$th atom, which depends on the position of the $i$th atom. A high-performance gate requires \cite{Gorshkov:07:cavity, Hao:15, Das:16} $C\gg 1$. While the single-atom cooperativity $g^2/\kappa\gamma= 0.14$ is below 1, $C$ can easily be made much larger than 1 because $N_a$ is large. Unless otherwise noted, all data in this paper are recorded with $C= 21(1)$, see appendix \ref{app-EIT}.

\subsection{Timing and Dual-Rail Setup}

\label{app-timing-and-dual-rail}

We prepare a new atomic sample every 4.3 s. Once a cold atomic ensemble has been prepared, we repeat the quantum gate experiment many times. Each repetition causes a tiny amount of photon-recoil heating and evaporative atom loss. To avoid large changes of $N_a$ and $T$, we discard the atomic ensemble after $10^4$ repetitions of the experiment and we prepare a new atomic ensemble. The EIT coupling light creates a repulsive potential for the atoms \cite{Baur:14, Tiarks:19}. To make its effect on the atomic density distribution negligible, each repetition of the experiment which lasts 3 $\mu$s is followed by 97 $\mu$s of hold time with only the ODT applied, yielding a repetition rate of 10 kHz. Including the time needed to prepare the atomic ensemble, the \emph{average} repetition rate is 2.3 kHz.

One quantum gate experiment consists of three steps. First, the control photon enters the setup; second, the target photon enters the setup and leaves it; and third, the delayed control photon leaves the setup. The dark time between consecutive steps is negligible. Photons which leave the setup are readily detected.

During step one, the leftmost PBS in Fig.\ \ref{fig-scheme}(a) converts the polarization qubit of the incoming control light pulse into a dual-rail qubit. One rail impinges upon the cavity and is stored as a stationary Rydberg excitation in state $|r'\rangle$ during the first step of the experiment. During step three, this photon is retrieved. The storage and retrieval results in a delay of $t_c+t_t$. The rightmost PBS in Fig.\ \ref{fig-scheme}(a) overlaps the retrieved control light with the control light from the other rail such that the dual-rail qubit is converted back into a polarization qubit. To achieve high fidelity of the gate, the components from both rails must have the same delay. Hence, control light in the bypass rail is delayed in a 400-m-long polarization-maintaining single-mode fiber, which causes a delay of 2.0 $\mu$s.

During step two, the polarization qubit of the incoming target pulse is also converted into a dual-rail qubit. One rail is reflected from the cavity. Upon this reflection, a conditional $\pi$ phase shift is acquired. A PBS overlaps the reflected target light with target light in the other rail, which simply bypasses the cavity. This overlap converts the dual-rail qubit back into a polarization qubit and simultaneously separates the target light from the counterpropagating control light. Two of the mirrors shown in Fig.\ \ref{fig-scheme}(a) are mounted on piezos, which are used to stabilize the path length differences between the rails for each qubit based on measurements with reference light. Quarter wave plates, omitted in Fig.\ \ref{fig-scheme}(a), map between the linear polarizations $H$ and $V$ and the circular polarizations required for driving the desired atomic transitions.

\section{Process Matrix Model}

\label{app-process-matrix}

Here, we model the main imperfections in the process matrix. We start from the process matrix $\chi^{\mathcal P}$ written with respect to the orthonormal basis $\mathcal P$, which consists of the matrices $|u_i\rangle\langle u_j|$, where $(|u_1\rangle, \linebreak[1] |u_2\rangle, \linebreak[1] |u_3\rangle, \linebreak[1] |u_4\rangle)= (|HH\rangle, \linebreak[1] |HV\rangle, \linebreak[1] |VH\rangle, \linebreak[1] |VV\rangle)$ denotes the CPHASE basis. Much like in Eq.\ \eqref{E-rho-chi}, $\chi^{\mathcal P}$ is defined by $\mathcal E(\rho)= \sum_{i,j,k,\ell=1}^{4} \linebreak[1] |u_i\rangle\langle u_j| \rho |u_\ell\rangle\langle u_k| \chi^{\mathcal P}_{i,j,k,\ell}$. Note that we use two indices to number the elements of the orthonormal basis $\mathcal P$. Hence, $\chi^{\mathcal P}$ has four indices running from 1 through 4.

\begin{figure}[!tb]
\centering
\includegraphics[width=\columnwidth]{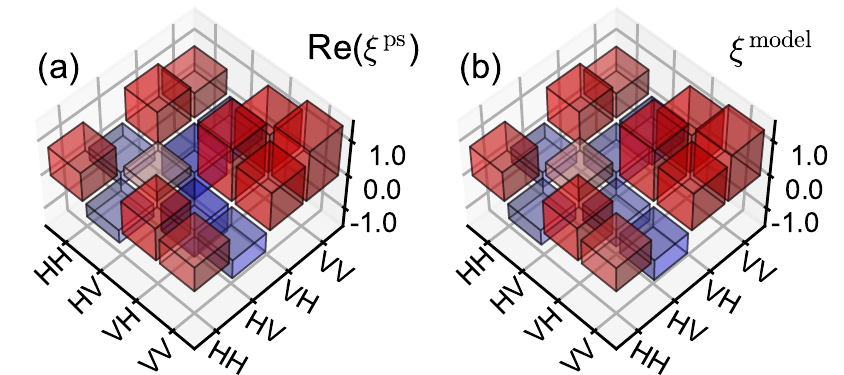}
\caption{(a) The real part $\Re(\xi^\text{ps})$ of the measured postselected reduced process matrix contains all the dominant experimental imperfections of the gate. (b) A model accounting for two single-qubit visibilities, which are used as free fit parameters, and for the fixed known efficiencies agrees well with the experimental data in (a).}
\label{fig-app-xi}
\end{figure}

The measured CPHASE truth table in Fig.\ \ref{fig-truth-tables}(a) is almost perfect. This implies $\langle u_i|\mathcal E(|u_j\rangle\langle u_j|)|u_i\rangle \approx 0$ if $i\neq j$, which is equivalent to $\chi^{\mathcal P}_{i,j,i,j}\approx \chi^{\mathcal P}_{i,i,i,i} \delta_{i,j}$. As in Ref.\ \cite{Nielsen:00}, we expect that $\mathcal E$ is a completely positive map, which implies that $\chi^{\mathcal P}$ is a positive matrix. Hence, $|\chi^{\mathcal P}_{i,j,k,\ell}|^2\leq \chi^{\mathcal P}_{i,j,i,j} \chi^{\mathcal P}_{k,\ell,k,\ell}\approx \chi^{\mathcal P}_{i,i,i,i} \delta_{i,j} \chi^{\mathcal P}_{k,k,k,k} \delta_{k,\ell}$, which implies
\begin{align}
\label{xi}
\chi^{\mathcal P}_{i,j,k,\ell}
\approx \xi_{i,k}\delta_{i,j} \delta_{k,\ell}
,&&
\xi_{i,k}
= \chi^{\mathcal P}_{i,i,k,k}
.\end{align}
Therefore, only 16 out of the 256 matrix elements of $\chi^{\mathcal P}$ are expected to deviate noticeably from zero. We find that the remaining 240 matrix elements of the measured $\chi^{\mathcal P}$ are indeed negligible. Hence, the relevant information about $\chi^{\mathcal P}$ is contained in the $4\times 4$ matrix $\xi$, which we call the reduced process matrix. The postselected version of $\xi$ is $\xi^\text{ps}= \xi/\bar\eta$. The real part of this is shown in Fig.\ \ref{fig-app-xi}(a). We expect and find experimentally that the imaginary parts of all matrix elements of $\xi$ are negligible. Hence, all the dominant imperfections of the gate are shown in Fig.\ \ref{fig-app-xi}(a). For the ideal CPHASE gate, one would obtain $\xi^\text{id}_{i,k}= (-1)^{\delta_{i,2}+\delta_{k,2}}$.

Somewhat similarly to Ref.\ \cite{Tiarks:19}, we now construct a model for the measured $\xi$ by assuming that each experimental shot is described by a linear map from state vectors onto state vectors defined by
\begin{subequations}
\begin{gather}
\label{beta-c-beta-t-a}%
|HH\rangle
\mapsto \sqrt{\eta_1} |HH\rangle
, \quad
|HV\rangle
\mapsto - e^{i\beta_t} \sqrt{\eta_2} |HV\rangle
, \\
\label{beta-c-beta-t-b}%
|VH\rangle
\mapsto e^{i\beta_c} \sqrt{\eta_3} |VH\rangle
, \quad
|VV\rangle
\mapsto e^{i(\beta_c+\beta_t)} \sqrt{\eta_4} |VV\rangle
,\end{gather}
\end{subequations}
where $(\eta_1, \linebreak[1] \eta_2, \linebreak[1] \eta_3, \linebreak[1] \eta_4)= (\eta_{HH}, \linebreak[1] \eta_{HV}, \linebreak[1] \eta_{VH}, \linebreak[1] \eta_{VV})$ and where $\beta_c$ and $\beta_t$ are real, single-qubit phases, which exhibit shot-to-shot fluctuations. We denote the average over these fluctuations as $\overline{\cdots}$. We assume that the random variables $\beta_c$ and $\beta_t$ are uncorrelated, we abbreviate the single-qubit visibilities $V_c= \overline{e^{i\beta_c}}$ and $V_t= \overline{e^{i\beta_t}}$, and we assume that $V_c$ and $V_t$ are real. We calculate a reduced postselected process matrix $\xi^\text{ps}$ from Eqs.\ \eqref{beta-c-beta-t-a} and \eqref{beta-c-beta-t-b}, average over the variables $\beta_c$ and $\beta_t$, and obtain
\begin{align}
\xi^\text{model}_{i,k}
= \overline{\xi^\text{ps}_{i,k}}
= (-1)^{\delta_{i,2}+\delta_{k,2}} V_{i,k} \frac{\sqrt{\eta_i\eta_k}}{\bar \eta}
\end{align}
with a visibility matrix
\begin{align}
V
=
\begin{pmatrix}
1 & V_t & V_c & V_cV_t \\
V_t & 1 & V_cV_t & V_c \\
V_c & V_cV_t & 1 & V_t \\
V_cV_t & V_c & V_t & 1 \\
\end{pmatrix}
=
\begin{pmatrix}
1 & V_c \\
V_c & 1 \\
\end{pmatrix}
\otimes
\begin{pmatrix}
1 & V_t \\
V_t & 1 \\
\end{pmatrix}
.\end{align}
We perform a least-squares fit of this model to the experimental data for $\xi^\text{ps}$ using the fixed, known values of the $\eta_i$ together with two free fit parameters $V_c$ and $V_t$. This yields the best-fit values quoted in Sec.\ \ref{sec-tomography}. Figure \ref{fig-app-xi}(b) shows the resulting $\xi$ from the model, which agrees well with the experimental data in Fig. \ref{fig-app-xi}(a).

The postselected process fidelity between the model and the ideal process is 78(2)\%, which agrees well with the postselected process fidelity of $F^\text{ps}_\text{pro}= 81(2)\%$ measured in the experiment. The contributions made by imperfect visibilities dominate because when setting the visibilities (efficiencies) as perfect in the model, the postselected process fidelity between the model and the ideal process becomes 95\% (82(2)\%).

In passing, we mention two points which follow from Eq.\ \eqref{xi}. First, combining Eq.\ \eqref{xi} with $\Theta= \sum_{i,j,k,\ell=1}^4
\linebreak[1]
|u_\ell\rangle \langle u_k|u_i\rangle \langle u_j| \chi^{\mathcal P}_{i,j,k,\ell}$ yields $\Theta \approx \sum_{i=1}^4 |u_i\rangle\langle u_i| \chi_{i,i}$, i.e.\ $\Theta$ is diagonal in the CPHASE basis. Second, for the input state $|DD\rangle$ which would ideally create the Bell state $|\psi_\text{Bell}\rangle= \frac12 \sum_{i=1}^4 (-1)^{\delta_{i,2}} |u_i\rangle$, Eq.\ \eqref{xi} yields a Bell-state fidelity $F_\text{Bell}= \langle \psi_\text{Bell}|\mathcal E(|DD\rangle\langle DD|)|\psi_\text{Bell}\rangle\approx \frac1{16} \sum_{i,k=1}^4
\linebreak[1]
(-1)^{\delta_{i,2}+\delta_{k,2}} \xi_{i,k}= \frac1{16} \tr(\xi^\text{id}\xi)= \frac1{16} \tr(\chi^{\mathcal P,\text{id}}\chi^{\mathcal P})$. The latter is the process fidelity $F_\text{pro}$ \cite{Gilchrist:05}.

\section{Experimental Limitations on the Efficiency}

\label{app-efficiency-limits}

Here, we discuss the main effects which limit the efficiency. The measured combined efficiency of storage and retrieval of the control photon is $\eta_\text{sr}= 39(1)$\%, in the absence of a target photon. This number is quite a bit lower than the theoretical expectation of 66\%, see appendix \ref{app-storage}. This discrepancy is probably caused by laser phase noise during storage and during retrieval. The measured transmission of the control photon through the delay fiber is $\eta_f= 65$\%, including mode matching. This is compatible with the specified attenuation of $-4.3$ dB/km, which would suggest 67\%.

The measured reflection coefficient for the target light is $|\mathcal R|^2= 90(1)$\% in the absence of a control pulse. However, if a control photon was previously stored in state $|r'\rangle$, this causes a Rydberg blockade and the reflectivity of the target light becomes $|\mathcal R_b|^2$. At the same time, the fact that a target photon is reflected from the cavity reduces the combined efficiency of storage and retrieval of the control photon to $\eta_{\text{sr},t}\leq \eta_\text{sr}$. Most of the experimental data in this paper are not sensitive to $\eta_{\text{sr},t}$ and $|\mathcal R_b|^2$ separately, but only to the product $\eta_{\text{sr},t} |\mathcal R_b|^2$. We measure $\eta_{\text{sr},t} |\mathcal R_b|^2= 17(2)$\%. If we naively assume $\eta_{\text{sr},t}= \eta_\text{sr}$, this would correspond to $|\mathcal R_b|^2= 43(4)$\%.

Note that $|\mathcal R|^2$ is nonideal partly because dephasing of the Rydberg state $|r\rangle$ causes the EIT transmission to be less than 100\% so that the resonator is not infinitely overcoupled. Conversely, $|\mathcal R_b|^2$ is nonideal partly because the finite atom number causes the absorption without EIT to be less than 100\% so that the resonator is not infinitely undercoupled. Note that $|\mathcal R_b|^2$ could be made much larger e.g.\ by increasing the cavity finesse. However, this would decrease $|\mathcal R|^2$. In our experiment, the parameters were chosen to maximize $\bar\eta$.

Combining the above numbers and ignoring possible loss if the target photon bypasses the cavity, one expects the efficiencies $(\eta_\text{sr}, \linebreak[1] \eta_{\text{sr},t} |\mathcal R_b|^2, \linebreak[1] \eta_f, \linebreak[1] \eta_f|\mathcal R|^2)= (39, \linebreak[1] 17, \linebreak[1] 65, \linebreak[1] 59)\%$ for the CPHASE basis states. The arithmetic mean thereof is 45\%. These values include only the main sources of imperfections, namely the delay fiber, the coefficients for reflection of the target photon from the cavity, and the efficiency of storage and retrieval. Beyond that there are trivial optical losses, e.g.\ on imperfect antireflection coatings on wave plates, lenses, PBSs, etc., some of which are not shown in Fig.\ \ref{fig-scheme}(a) for clarity. Taking those trivial losses into account, we obtain the measured efficiencies quoted in Sec.\ \ref{sec-efficiency}.

\section{Coincidence Rate}

\label{app-coincidence}

The experimental apparatus includes two identical detection setups, one at each output port of the gate. To perform a polarization resolved detection of signal photons, each detection setup consists of wave plates and one PBS. Each of the two output ports of each PBS is sent to a superconducting nanowire single-photon detector (SNSPD). Polarization-maintaining single-mode fibers are used to transport the light to the SNSPDs. The fiber coupling efficiency of approximately 85\% is caused partly by mode matching, partly by losses in the input-output couplers of the fiber, and partly by losses caused by a fiber splice. The quantum efficiency of each SNSPD is specified to be above 90\%. Hence, if one control and one target photon leave the gate, the probability of detecting both is approximately $(0.85\times 0.90)^2= 0.59$. Multiplying this by the 41.7\% average efficiency of the gate, one obtains a 24\% probability of detecting a two-photon coincidence per incoming photon pair. Multiplying this with the \emph{average} repetition rate from appendix \ref{app-timing-and-dual-rail} yields a two-photon coincidence rate of 560 s$^{-1}$ per incoming photon pair. This is a factor of $1.3\times 10^4$ larger than in our previous apparatus \cite{Tiarks:19} for reasons discussed in appendix \ref{app-improved-coincidence}. This is a tremendous improvement.

The mean photon number $\bar n_c$ ($\bar n_t$) in the incoming control (target) pulse should not be too large because that would degrade the measured postselected fidelity due to events with more than one incoming control (target) photon. On the other hand, $\bar n_c$ and $\bar n_t$ should not be too small because that would increase the data acquisition time and make the system more sensitive to dark counts. The data in Figs.\ \ref{fig-truth-tables} and \ref{fig-process-tomography} are measured with $\bar n_c= 0.14$ and $\bar n_t= 0.13$. We multiply $\bar n_c \bar n_t$ by the two-photon coincidence rate per incoming photon pair and obtain a naive estimate for the two-photon coincidence rate of 10 s$^{-1}$. In daily alignment, we measure 9 s$^{-1}$. It is this high coincidence rate which makes it possible to perform demanding measurements in a reasonable data acquisition time. For the process tomography in Fig.\ \ref{fig-process-tomography}, for example, the data acquisition time was 69 min.

\section{GHZ States}

\label{app-GHZ}

For simplicity, we fit $S_\vartheta^{(N)}= p\cos(N\vartheta)$ with a free fit parameter $p\in [0,1]$ to the data in Fig.\ \ref{fig-GHZ}. This simple curve is expected, e.g.\ if $\rho_\text{out}= p|\psi_N\rangle\langle\psi_N| +(1-p)I$, where $I$ is the identity matrix. As an alternative to the direct calculation of $\mathcal C_N$ as described above, one can insert the fit curve into this calculation, which yields $\mathcal C_N= p$ and makes use of all the available data instead of only the points for $\vartheta= \pi k/N$ with $k\in\{0,1,2,\dots,N-1\}$. We use this method only for processing the $N=6$ data.

\begin{figure}[!tb]
\centering
\includegraphics[width=\columnwidth]{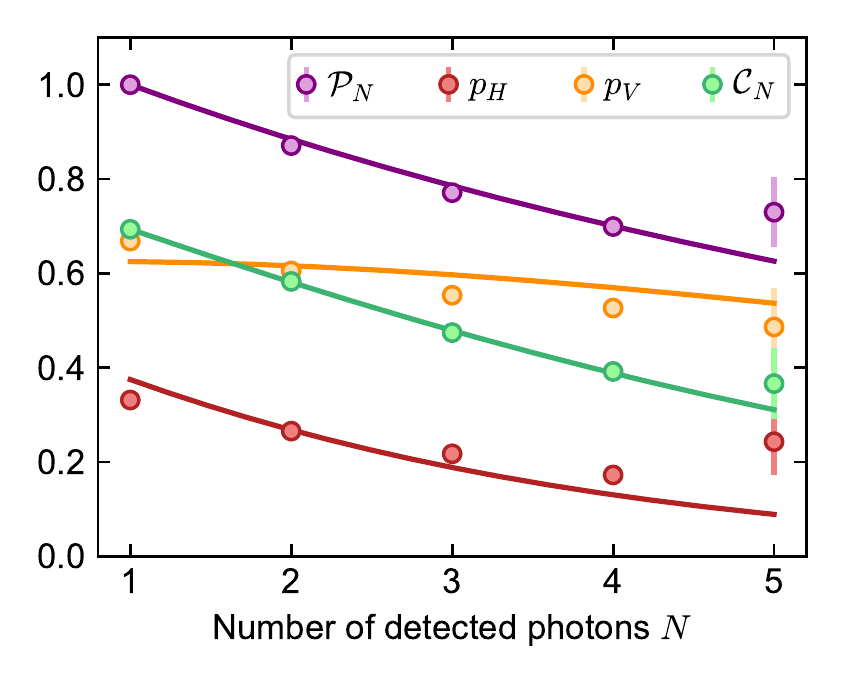}
\caption{The populations $p_H$, $p_V$, and $\mathcal P_N$ and the coherence $\mathcal C_N$ of the observed GHZ states depend on the number of detected photons $N$. The lines show results from a simple model, which contains only one free fit parameter $V_c$. This parameter affects only $\mathcal C_N$. The model agrees fairly well with the data. According to the model, the nonideal visibilities are the dominant experimental imperfections.}
\label{fig-app-GHZ}
\end{figure}

The measurements for $N\leq 5$ use $\bar n_c=0.31$ and $\bar n_t=0.41$. As a matter of fact, all $N\leq 5$ data are extracted from the same raw data by postselecting upon different numbers of detected photons. The measurement for $N=6$ is from a different data set, using $\bar n_c= 0.28$ and $\bar n_t= 0.88$.

To understand the present limitations of the GHZ-state production, we consider Fig.\ \ref{fig-app-GHZ} which shows how $p_H$, $p_V$, $\mathcal P_N$, and $\mathcal C_N$ depend on $N$. The lines represent a simple model, see appendix \ref{app-GHZ-model}
\begin{gather}
\label{p-H}
p_H
= \frac1{\eta_N} \frac{\eta_\text{sr}}{2} \left(\frac{1+|\mathcal R_b|^2+ 2V_t \Re(-\mathcal R_b)}4 \right)^{N-1}
,\\
p_V
= \frac1{\eta_N} \frac{\eta_f}{2} \left(\frac{1+|\mathcal R|^2+ 2V_t \Re(\mathcal R)}4 \right)^{N-1}
,\\
\label{eta-N}
\eta_N
= \frac{\eta_\text{sr}}2 \left(\frac{1+|\mathcal R_b|^2}2\right)^{N-1}
+ \frac{\eta_f}2 \left(\frac{1+|\mathcal R|^2}2\right)^{N-1}
,\\
\label{C-N}
\mathcal C_N
= \frac{V_c}{\eta_N} \frac{\sqrt{\eta_\text{sr}\eta_f}}{2} \left(\frac{1  - \mathcal R_b \mathcal R^* + V_t ( \mathcal R^*-\mathcal R_b )}4\right)^{N-1} +
\!
\text{c.c}
.\end{gather}
Here, $\eta_N$ is the probability that no photon is lost for an input state with exactly one control photon and exactly $N-1$ target photons. For simplicity, we assume $\mathcal R> 0$ and $\mathcal R_b< 0$.

For $p_H$, $p_V$, and $\mathcal P_N= p_H+p_V$, the model uses the target visibility $V_t$ from appendix \ref{app-process-matrix} together with the efficiencies expressed by the quantities $\eta_\text{sr}$, $\eta_f$, $|\mathcal R_b|^2$, and $|\mathcal R|^2$ from appendix \ref{app-efficiency-limits} and it has no free fit parameters. For $\mathcal C_N$, the model additionally uses one free fit parameter $V_c$ with the best-fit value $V_c= 72\%$. Note that Eqs.\ \eqref{p-H}--\eqref{eta-N} are independent of $V_c$. The model agrees fairly well with the experimental data.

The best-fit value $V_c= 72\%$ obtained here is lower than the value $V_c= 86(4)\%$ extracted from the process matrix in appendix \ref{app-process-matrix}. A detailed analysis shows that this is mostly due to the incoming mean photon numbers $\bar n_c$ and $\bar n_t$, which are larger here than in appendix \ref{app-process-matrix}. In essence, the nonideal detection efficiency combined with the Poisson distributions of the incoming photon numbers reduces $\mathcal C_N$ by a constant factor, which is independent of $N$. When varying $V_c$ to fit Eq.\ \eqref{C-N} to the experimental data, this causes an effective reduction of the best-fit value of $V_c$.

The combination of the nonideal detection efficiency with the Poisson distributions of the incoming photon numbers also contributes a factor of 0.98 to the postselected process fidelity $F^\text{ps}_\text{pro}$ extracted from Fig.\ \ref{fig-process-tomography}.

Assuming ideal efficiencies in the model has little effect on $\mathcal P_N$ and only a moderate effect on $\mathcal C_N$. Hence, much like in appendix \ref{app-process-matrix}, we conclude that the nonideal visibilities are the dominant experimental imperfections.

\begin{table}[t!]
\centering
\begin{tabular*}{\columnwidth}{c@{\extracolsep\fill}ccccc}
\hline \hline
$N$ & 1 & 2 & 3 & 4 & 5 \\
$R_N$ (s$^{-1}$) & 171(1) & 46.0(2) & 6.0(1) & 0.46(2) & 0.028(5) \\
\hline \hline
\end{tabular*}
\caption{The $N$-fold coincidence rate $R_N$ decreases with increasing $N$.}
\label{tab-coincidence-rates}

\end{table}

The rates $R_N$ at which $N$-fold coincidences are detected are listed in table \ref{tab-coincidence-rates}. Note that $R_N$ describes the coincidence events with exactly one detected control photon and exactly $N-1$ detected target photons. Here, $R_N$ is based on the \emph{average} repetition rate, which includes the time needed to prepare the atomic ensemble.

To develop a simple model for the rates $R_N$, we approximate $\eta_c= \frac12(\eta_\text{sr}+\eta_f)$ and $\eta_t= \frac14(2+|\mathcal R|^2+|\mathcal R_b|^2)$ as uncorrelated mean efficiencies with which each control and target photon is transmitted. We use the efficiency from appendix \ref{app-coincidence} with which each photon is detected as $\eta_d= 0.85\times 0.90$. We obtain a rough estimate for the mean number of detected control (target) photons of $\nu_c= \eta_d \eta_c \bar n_c= 0.12$ ($\nu_t= \eta_d \eta_t \bar n_t= 0.26$.) We assume uncorrelated Poisson distributions for the probabilities of detecting a given number of control and target photons. We obtain the probability $p_N= \nu_c e^{-\nu_c} \frac1{(N-1)!} \nu_t^{N-1} e^{-\nu_t}$ of detecting exactly one control photon and exactly $N-1$ target photons. Multiplying this with the average repetition rate from appendix \ref{app-timing-and-dual-rail}, we find fairly good agreement with the experimental data in table \ref{tab-coincidence-rates}.

\section{Cooling and Trapping of Atoms}

\label{app-cooling}

To prepare the atomic ensemble in our new vacuum system, a two-dimensional magneto-optical trap (MOT) generates a continuous cold beam of $^{87}$Rb atoms, which is captured in a three-dimensional (3D) MOT. The latter is overlapped with the crossed-beam ODT with a light wavelength of 1064 nm. One ODT beam has a power of 0.45 W, a horizontal wave vector, and beam waists of $w_y=15$ $\mu$m and $w_z=11$ $\mu$m. The other ODT beam has a power of 0.10 W, a vertical wave vector, and beam waists of $w_x= w_z=12.5$ $\mu$m. After collecting atoms in the 3D MOT for 0.5 s, the parameters of the 3D MOT are altered similarly to Ref.\ \cite{Kuppens:00} to increase the atom number in the ODT. In particular, the power of the repumping laser is much reduced. After 30 ms, both MOTs are switched off and polarization gradient cooling is applied for 50 ms. Then, the atoms are held in the ODT.

The atomic ensemble in the ODT is further cooled by applying the first two stages of Raman cooling described in Ref.\ \cite{Urvoy:19} but in a crossed-beam ODT. This Raman cooling lasts 0.5 s and it spin polarizes the atoms by optically pumping them into the state $|g\rangle$. During the cooling, the powers of both ODT beams are progressively lowered. For a cold dilute ensemble in a deep ODT, we measured a $1/e$ lifetime of the atom number of 30 s, which sets a limit to background gas collisions. After the Raman cooling, the power of the ODT is lowered to 4.5 mW (5 mW) for the ODT beam with a horizontal (vertical) wave vector to reach $T=0.44$ $\mu$K and $N_a= 260$.

This corresponds to a peak density of $\varrho_\text{peak}= 6 \times 10^{11}$ cm$^{-3}$ and a peak phase-space density of $0.014$. Note that density-dependent dephasing resulting from collisions of a Rydberg atom with surrounding ground-state atoms is less of an issue than it was in Ref.\ \cite{Baur:14} because we operate at a much lower principal quantum number. For signal light at $\lambda=780$ nm driving a cycling transition, the cross section for absorption of a photon is $\sigma_\text{cyc}= 3\lambda^2/2\pi$ and the maximal on-axis ($x=y=0$) on-resonance optical depth is $d_{t,\text{peak}}= \sigma_\text{cyc} \int dz \varrho= 0.8$.

\section{Improved Coincidence Rate}

\label{app-improved-coincidence}

Here, we discuss the physical origin of the factor of $1.3\times10^4$ improvement of the coincidence rate per incoming photon pair compared to our previous apparatus \cite{Tiarks:19}, in which the two-photon coincidence rate was $1.3\times10^{-5} \times 0.56\text{ kHz} /(0.33\times0.50)= 0.044$ s$^{-1}$ per incoming photon pair. An obvious contribution comes from the higher average repetition rate, which was $10^4/(18 \text{ s})=0.56$ kHz in Ref.\ \cite{Tiarks:19}. Another contribution comes from the higher efficiency of collecting and detecting an outgoing signal photon, which was 25\% in Ref.\ \cite{Tiarks:16}, caused partly by the different fiber coupling efficiency and partly by the different efficiency of the single-photon detector, which was 50\% in Ref.\ \cite{Tiarks:19}. Yet another contribution comes from the higher average efficiency of the gate, which in Ref.\ \cite{Tiarks:19} included only loss in the atomic ensemble and was $(7.7,2.3,1.5,0.45)\%$ for the CPHASE basis states with an average of 3.0\%. When creating a Bell state, the polarizations with higher gate efficiency were attenuated in the first interferometer in Ref.\ \cite{Tiarks:19} before the light impinged on the atoms. On the upside, these imbalanced input intensities resulted in well-balanced output intensities, which boosted the postselected Bell-state fidelity. On the downside, for this input state the efficiency is the harmonic mean of the CPHASE efficiencies $(\frac14 \sum_{i=1}^4 \eta_i^{-1})^{-1}= 1.2\%$, quite a bit less than 3.0\%. Such imbalanced input intensities are not used in our present work. The contributions listed so far explain a factor of $\frac{2.3\text{ kHz}}{0.56\text{ kHz}} \left(\frac{0.85\times 0.90}{0.25}\right)^2 \frac{41.7\%}{1.2\%}= 1.4\times 10^3$, which is the biggest part of the overall effect.

The remaining factor of approximately 9 comes from three contributions. First, the second interferometer, which was needed to remove a frequency shift, contributed to loss of photons between the atomic ensemble and the detector in Ref.\ \cite{Tiarks:19}. Second, the light was coupled into a fiber in front of this interferometer and into a second fiber behind this interferometer so that two fiber coupling efficiencies occurred between the atoms and the detector. This interferometer is no longer needed in the new apparatus and now light passes through only one fiber between the atoms and the detector. Third, the data acquisition time for quantum-state tomography of one Bell state in Ref.\ \cite{Tiarks:19} was long. These data were accumulated from several nonconsecutive days. Some components were not perfectly kept at optimal performance over this full time span.

For comparison, the single-atom experiment in Ref.\ \cite{Hacker:16} measured a two-photon coincidence rate of 0.033 s$^{-1}$ in daily alignment. Division by $\bar n_c\bar n_t$ with $\bar n_c= \bar n_t= 0.17$ yields a two-photon coincidence rate of 1.1 s$^{-1}$ per incoming photon pair. This is a factor of 25 higher than in our previous experiment \cite{Tiarks:19} and a factor of 520 lower than our present experiment.

\section{Cavity EIT}

\label{app-EIT}

\subsection{Theory}

The one-sided cavity is characterized by the complex reflection coefficient $r_\text{in}$ of the input-output (I/O) coupler and by the product $r_H$ of the complex reflection coefficients of all three highly reflective (HR) cavity mirrors. We use these quantities to define
\begin{align}
\mathcal F_\text{in}
&
= \frac{\Delta\omega_\text{ax}}{2\kappa_\text{in}}
= \frac{-\pi}{\ln|r_\text{in}|}
,\\
\mathcal F_H
&
= \frac{\Delta\omega_\text{ax}}{2\kappa_H}
= \frac{-\pi}{\ln|r_H|}
,\\
\mathcal F
&
= \frac{\Delta\omega_\text{ax}}{2\kappa}
= \frac{-\pi}{\ln |r_\text{in}r_H|}
.\end{align}
Here, the decay rate coefficient $\kappa_\text{in}$ ($\kappa_H$) describes the decay of the intracavity field caused by the I/O coupler (HR mirrors), $\kappa= \kappa_\text{in}+\kappa_H$ is the total decay rate, and $\mathcal F$ is the empty-cavity (i.e.\ $N_a=0$) finesse. For simplicity, the I/O coupler is assumed to be lossless.

In the absence of a Rydberg blockade, the reflection of the signal light field impinging on the cavity is described by the complex coefficient \cite{Das:16}
\begin{align}
\label{R}
\mathcal R
= -1+\frac{2\kappa_\text{in}}{\kappa(1+C_\text{eff})-i\Delta_c}
,\end{align}
where $\Delta_c= \omega-\omega_c$ is the detuning of the angular frequency $\omega$ of the signal light from the nearest cavity resonance $\omega_c$. In addition, we abbreviate an effective cooperativity
\begin{align}
\label{C-eff}
C_\text{eff}
= C \frac{\Gamma_e}{\Gamma_e-2i\Delta_s +\frac{|\Omega|^2}{\gamma_{rg}-2i(\Delta_\text{co}+\Delta_s)}}
,\end{align}
where $1/\Gamma_e= 1/2\gamma$ is the $1/e$ lifetime for population decay of state $|e\rangle$ and where $\gamma_{rg}$ describes the decay of the density matrix element $\rho_{rg}$ according to $(\partial_t +\frac12\gamma_{rg}) \rho_{rg}= 0$, which is typically dominated by dephasing \cite{Schmidt-Eberle:20}. Note that $\Delta_s= \omega-\omega_{ge}$ ($\Delta_\text{co}= \omega_\text{co}-\omega_{er}$) is the detuning of the angular frequency $\omega$ ($\omega_\text{co})$ of the EIT signal (coupling) light from the atomic resonance $\omega_{ge}$ ($\omega_{er}$), and $\Omega$ is the Rabi frequency of the EIT coupling light.

The model of Eqs.\ \eqref{R} and \eqref{C-eff} relies on neglecting all but one of the axial cavity modes. This is a good approximation, if $\mathcal F\gg 1$, $|\Delta_c|\ll \Delta\omega_\text{ax}$, and $|C_\text{eff}|\ll \mathcal F/\pi$. The last condition is equivalent to saying that the transversely averaged optical depth in a single round-trip in the cavity must be small. In addition, Eq.\ \eqref{R} assumes perfect matching of the incoming light to the fundamental transverse mode of the cavity. Moreover, if the ensemble is small in the transverse direction, this might result in a fairly large on-axis optical depth $d_{t,\text{peak}}$. Hence, some part of the atoms might experience a noticeable shadow cast by other atoms upon a single passage of the light through the medium. The model of Eqs.\ \eqref{R} and \eqref{C-eff} with $C= \sum_{i=1}^{N_a} |g_i|^2/\kappa\gamma$ does not take into account a possible transverse inhomogeneity of such a shadow effect.

\subsection{Experiment}

With a standard method, based on light transmitted through the cavity, we extract a mode matching of 99.1\% of the incoming light to the fundamental transverse cavity mode. With an alternative method, based on light reflected from the cavity, we extract a mode matching of 98.0\%.

Measurements of $\Delta\omega_\text{ax}$ and $\kappa$ yield $\mathcal F= 350$. Combining this with the measured resonant empty-cavity reflection coefficient $|\mathcal R_\text{empty}|^2= 0.932$ and with Eq.\ \eqref{R} with $C_\text{eff}= 0$ yields $\mathcal F_H= 2.0\times10^4$. From this result, we obtain $|r_\text{in}|^2= 98.3\%$ for the I/O coupler and $|r_H|^{2/3}= 99.990\%$ for each of the three HR mirrors. The latter is consistent with the finesse measured in a test resonator made of four HR mirrors from the same coating run.

\begin{figure}[!tb]
\centering
\includegraphics[width=\columnwidth]{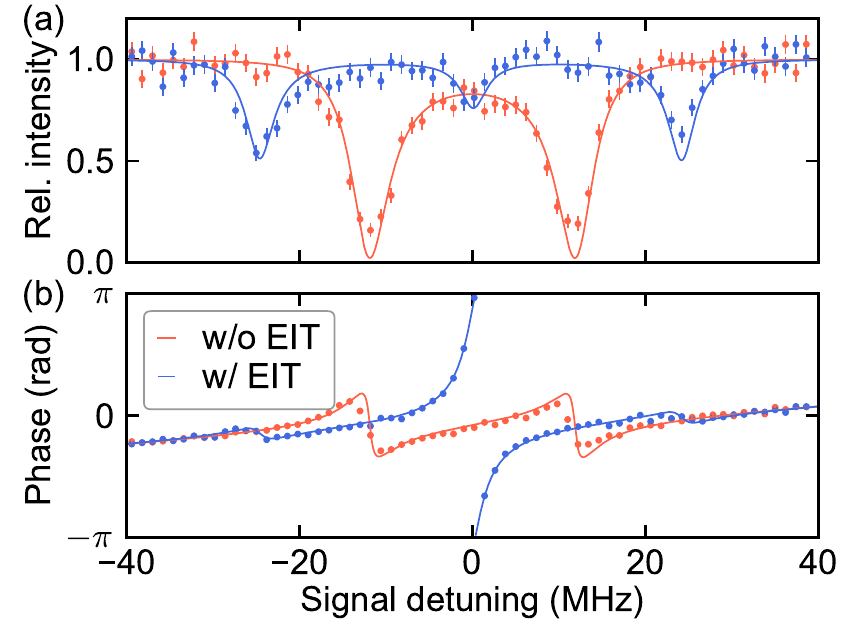}
\caption{Cavity Rydberg EIT spectra. (a) Relative intensity and (b) phase of the light reflected from the cavity shown as a function of the signal detuning $\Delta_c/2\pi$. Red (blue) data were recorded in the absence (presence) of EIT coupling light. The experiment is well within the regime of large normal-mode splitting. The blue data clearly show an EIT feature at zero detuning. Fits (lines) to the data (dots) reveal important experimental parameters, namely the collective cooperativity $C$, the EIT coupling Rabi frequency $\Omega$, and the coherence time $1/\gamma_{rg}$.}
\label{fig-spectra}
\end{figure}

For calibration purposes, we perform measurements without the control photon. Figures \ref{fig-spectra}(a) and \ref{fig-spectra}(b) show the relative intensity and the phase of target signal light reflected from the cavity as a function of the detuning $\Delta_c/2\pi$ of the signal light from the cavity resonance. The measurement is performed with $\Delta_\text{co}\approx 0$, and with $\omega_c\approx \omega_{ge}$, so $\Delta_c\approx \Delta_s$, see appendix \ref{app-geometry}. The red data are taken in the absence of EIT coupling light, $\Omega= 0$. The intensity clearly exhibits a normal-mode splitting, which is much larger than the linewidth. The phase shows two dispersive features with small amplitude, one at each normal mode. The blue data are taken in the presence of EIT coupling light. The resulting cavity EIT causes the intensity to exhibit a narrow EIT feature centered at $\Delta_c= 0$. In addition, the normal-mode splitting is somewhat increased. When tuning $\Delta_c$ through zero, cavity EIT causes a $2\pi$ phase boost.

At resonance, $\Delta_c= 0$, the phase is 0 ($\pi$) in the absence (presence) of EIT coupling light. This shows the origin of the conditional $\pi$ phase shift on which the gate is built, much like in Ref.\ \cite{Tiarks:16}. This is because if a stationary Rydberg excitation is added, then the Rydberg blockade will detune the EIT coupling light so far that the presence of the EIT coupling light becomes irrelevant. Hence, the presence (absence) of a stationary Rydberg excitation is expected to be equivalent to the absence (presence) of EIT coupling light.

To characterize the parameter values of the cavity Rydberg EIT system, we fit the above analytic model in Eqs.\ \eqref{R} and \eqref{C-eff} simultaneously to the data in Figs.\ \ref{fig-spectra}(a) and \ref{fig-spectra}(b). The fit to the red data in Fig.\ \ref{fig-spectra} yields the collective cooperativity $C= 21.4(2)$, where the values of $\mathcal F$, $\kappa$, and $\kappa_\text{in}$ are taken from previously described measurements and the value of the atomic linewidth $\Gamma_e/2\pi= 6.0666$ MHz is taken from the literature. A subsequent fit to the blue data yields the EIT coupling Rabi frequency $\Omega/2\pi= 42.7(3)$ MHz and the coherence time $1/\gamma_{rg}= 0.22(5)$ $\mu$s, where the previous values of $\mathcal F$, $\kappa$, $\kappa_\text{in}$, $\Gamma_e$, and $C$ are used.

Apart from $\bar n_c= 0$, $\bar n_t= 0.19$, and scanning $\omega$, this measurement is performed at the same settings as for the process tomography in Fig.\ \ref{fig-process-tomography}, e.g.\ with the same target pulse shape, the same target pulse duration, the same Rydberg state $|r\rangle= |50S_{1/2}, \linebreak[1] F{=}2, \linebreak[1] m_F{=}{-}2\rangle$, and the same electric field of 0.7 V/cm.

If $N_a= 260$ atoms were maximally coupled, then we would expect to measure $C= 37$. The discrepancy from the experimental observation is mostly due to the nonzero transverse sample size which reduces the expectation for $C$ by a factor of $[1+(2\sigma_x/w_c)^2]^{-1/2} \times \linebreak[1] [1+(2\sigma_y/w_c)^2]^{-1/2}= 0.54$ to $C= 20$. Including the transversely inhomogeneous shadow, discussed below Eq.\ \eqref{C-eff}, reduces this expectation to $C= 19$. Additional contributions might come from imperfect alignment of the cloud center relative to the cavity mode and uncertainties in the calibration of the measurement of the atom number based on absorption imaging. The overall factor resulting from these additional contributions seems to be small.

\section{Modeling Storage and Retrieval}

\label{app-storage}

In the adiabatic limit of slowly varying pulse envelopes, the combined efficiency of storage and retrieval can be modeled analytically, yielding \cite{Gorshkov:07:cavity}
\begin{align}
\label{eta-sr}
\eta_\text{sr}
= \left( \frac{\kappa_\text{in}}{\kappa} \; \frac{C}{C+1}\right)^2 e^{-\gamma_{rg}(t_c+t_\text{dark})}
,\end{align}
where $t_\text{dark}$ is the dark time between the end of storage and the beginning of retrieval. The atoms are not truly in the dark during this so-called dark time because the crossed-beam ODT is always on continuously in our present work. We vary $t_\text{dark}$ and measure the corresponding value of $\eta_\text{sr}$. This yields an approximately Gaussian decay with a $1/e$ time of 7 $\mu$s, which is fairly close to the expectation inside the ODT based on a minor extension of the model of Ref.\ \cite{Schmidt-Eberle:20}. Following Ref.\ \cite{Gorshkov:07:cavity}, we model this as an exponential decay and, for simplicity, we choose the same $1/e$ time $1/\gamma_{rg}= 7$ $\mu$s. Combining this with $C= 21$ and $t_c+t_\text{dark}= 2.0$ $\mu$s, Eq.\ \eqref{eta-sr} predicts $\eta_\text{sr}= 66.1$\%. In the limit $\gamma_{rg}\to 0$, Eq.\ \eqref{eta-sr} would predict $\eta_\text{sr}= 88.0$\%. Hence, according to the model, $\gamma_{rg}$ gives rise to more than half of the imperfections in $\eta_\text{sr}$. To test the quality of the adiabatic approximation used to derive Eq.\ \eqref{eta-sr}, we numerically solve the differential equations in Ref.\ \cite{Gorshkov:07:cavity} without approximations. This yields $\eta_\text{sr}= 64.7$\%. We conclude that the adiabatic approximation works well in this parameter regime.

Note that, for simplicity, $\gamma_{rg}$ is assumed to be time independent when deriving Eq.\ \eqref{eta-sr}. This is not necessarily realistic because issues like laser phase noise or fluctuations of the ambient magnetic or electric field can effectively increase \cite{Gea-Banacloche:95} $\gamma_{rg}$ during storage and retrieval, while they do not contribute to the decay of $\eta_\text{sr}$ as a function of the dark time \cite{Schmidt-Eberle:20}.

Reflecting one target photon from the cavity during the dark time decreases the combined efficiency of storage and retrieval from $\eta_\text{sr}$ to $\eta_{\text{sr},t}$. Measuring $\eta_{\text{sr},t}$ is nontrivial because we do not use a single-photon source for the target photon. Hence, to learn something about $\eta_{\text{sr},t}$ we rely on postselecting data upon detection of a reflected target photon. Essentially, this means that we measure a coincidence rate which reveals the product $\eta_{\text{sr},t}|\mathcal R_b|^2$ but not $\eta_{\text{sr},t}$ and $|\mathcal R_b|^2$ separately. Likewise, when aiming at measuring $|\mathcal R_b|^2$, we do not use a single-photon source for the control photon. Hence, we rely on postselecting data upon detection of a retrieved control photon. Essentially, this means that we measure the same coincidence rate as above.

\section{Polarization Conventions}

All the input and output polarizations refer to the polarization at the points at which the qubits enter and leave the gate, which is the part of the setup shown in Fig.\ \ref{fig-scheme}(a). For a plane light wave with wave vector $\bm k$, the electric field is given by $\bm E(\bm x,t)= \frac12 E_0 \bm u e^{i\bm k\cdot\bm x-i\omega t}+\text{c.c.}$, where $E_0$ is a complex amplitude, $\omega> 0$ is the angular frequency, and $\bm u$ is the complex polarization unit vector. If $\bm k$ is along the positive $z$ axis, then for horizontal (vertical) polarization $\bm u= \bm e_x$ ($\bm u= \bm e_y$), where $(\bm e_x, \bm e_y, \bm e_z)$ is the right-handed Cartesian basis. The polarization bases used here are connected by $|D\rangle= (|H\rangle+|V\rangle)/\sqrt2$, $|A\rangle= (|H\rangle-|V\rangle)/\sqrt2$, $|R\rangle= (|H\rangle-i|V\rangle)/\sqrt2$, and $|L\rangle= (|H\rangle+i|V\rangle)/\sqrt2$, where $R$ and $L$ denote right-hand and left-hand circular polarizations. These conventions for $R$ and $L$ are consistent with Ref.\ \cite{Jones:48}.

\section{Cavity Geometry}

\label{app-geometry}

We choose a ring resonator to minimize dephasing resulting from the combination of photon recoil and thermal atomic motion \cite{Schmidt-Eberle:20}. To achieve this minimization, the signal light and coupling light in Rydberg EIT must counterpropagate with each other. In a standing-wave resonator, the two traveling-wave components of the signal light standing wave could never simultaneously counterpropagate with the coupling light. Hence, a standing-wave resonator would lead to fast dephasing of at least one of the traveling-wave components. A ring resonator solves this problem.

Using a ring resonator brings the additional advantage that one can use the novel scheme shown in Fig.\ \ref{fig-scheme}(a) for converting the conditional $\pi$ phase shift into a two-photon gate for polarization qubits. In a one-sided standing-wave resonator, however, the light reflected from the cavity would exactly counterpropagate with the impinging light, thus requiring an additional effort to separate the beams.

In ring resonators, light is reflected from curved mirrors at a nonzero angle of incidence. This causes the resonator mode to develop an undesired astigmatism. To mitigate this problem, it is advantageous to use small angles of incidence which is possible, e.g., in a four-mirror bow-tie geometry.

In addition, the vacuum Rabi frequency $g$ for one maximally coupled atom should not be too small. In principle, one could compensate for small $g$ by increasing the atom number $N_a$ or by decreasing the cavity linewidth $\kappa$. However, in order to maintain a superatom geometry, the size of the atomic ensemble should be smaller than the blockade radius and the atomic density $\varrho$ should not be too large because otherwise collisions between a Rydberg atom and the surrounding ground-state atoms would result in density-dependent dephasing \cite{Baur:14}. Reducing $\kappa$ is not desirable either because that would make the experiment more susceptible to laser phase noise and to pulse bandwidth effects, which both deteriorate the postselected fidelity.

We note that $g_i= d_{eg} v(\bm x_i) \sqrt{\omega/2\hbar\epsilon_0}$, where $d_{eg}$ is the electric dipole moment on the $|g\rangle \leftrightarrow |e\rangle$ transition, $\bm x_i$ is the position of the $i$th atom, and $v(\bm x)$ is the cavity mode function, normalized to $\int d^3x |v(\bm x)|^2= 1$. In a traveling-wave cavity $|v(0)|= \sqrt{2/\pi w_c^2L_c}$, where the coordinate origin is at the cavity mode waist and $L_c= 2\pi c/\Delta\omega_\text{ax}$ is the cavity round-trip length. Hence, a short cavity with a small waist is desirable.

To realize a short bow-tie cavity with a small waist, one needs two concave mirrors with small radii of curvature not too far from the waist together with two convex mirrors, much like in Ref.\ \cite{Jia:18}. By varying the radii of curvature of the convex mirrors, one can also vary the frequency splittings to higher transverse modes, thus avoiding degenerate modes, which might lower the cavity finesse, a problem often encountered in near-concentric Fabry-P\'{e}rot resonators \cite{Haase:06,Turnbaugh:21}.

If the cavity mirrors had a non-negligible reflectivity for the 480-nm EIT coupling light, then one would need a triply resonant cavity, which is resonant with the signal light, the control coupling light, and the target coupling light. We evade this problem by choosing dielectric coatings with small reflectivities at 480 nm.

\begin{figure}[!tb]
\centering
\includegraphics[width=\columnwidth]{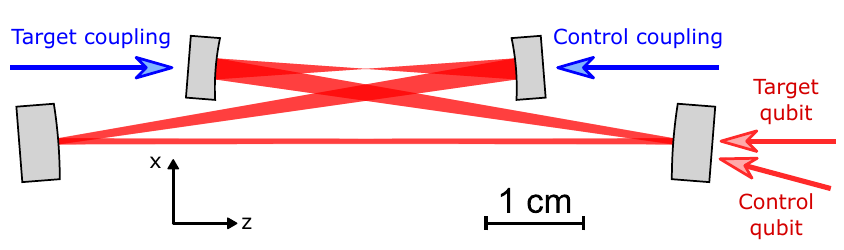}
\caption{Geometry of the bow-tie cavity. Two of the beams inside the cavity are parallel. These beams have lengths of $L_1= 30.75$ mm and $L_2= 63.00$ mm. The angle of incidence is 4.5$^\circ$. The two mirrors separated by the shorter distance $L_1$ are concave with radii of curvature of $-25$ mm. The cavity waist $w_c=8.5$ $\mu$m and the atomic ensemble are located midway between these two mirrors. The other two mirrors are convex with radii of curvature of 50 mm. One of the convex mirrors serves as the input-output coupler for signal light. The substrate of each concave (convex) mirror has a diameter of 6.35 mm (7.75 mm).}
\label{fig-geometry}
\end{figure}

The bow-tie geometry of the cavity is shown in Fig.\ \ref{fig-geometry}. The geometry and the dimensions resemble the setup in Ref.\ \cite{Jia:18}. For fixed $L_2$ and fixed angle of incidence, the resonator is stable for 30.34 mm $\leq L_1 \leq$ 31.24 mm. If the Rydberg atoms were too close to a surface, then that might cause rapid dephasing of the Rydberg states. The geometry chosen here places the nearest surface approximately 15 mm from the atomic ensemble which seems to be safe \cite{Jia:18}.

The EIT signal light drives the $|g\rangle \leftrightarrow |e\rangle$ transition, which requires $\sigma^-$ polarized light. The polarization convention for $\sigma^-$ is relative to the positive $z$ axis (see Fig.\ \ref{fig-geometry}) and essentially represents the direction of the atomic spins. The polarization convention for $L$ and $R$, however, is relative to the wave vector of the light. As the control (target) signal light has a wave vector parallel (antiparallel) to the positive $z$ axis, it is $R$ ($L$) polarized.

One can show that in the absence of birefringence the resonance frequency of the $L$ polarized cavity mode is independent of the propagation direction of the light. The same applies to the $R$ polarized cavity mode. As the two counterpropagating EIT signal light beams for control and target pulses should both be resonant with the atomic transition and with the cavity, the $L$ and $R$ polarized cavity modes should have negligible frequency splitting. Otherwise it is impossible to make both cavity modes resonant with the same atomic resonance. Among other things, this requires the cavity to be planar, i.e.\ that all the wave vectors of the cavity mode lie in one plane. This is a nontrivial condition for the fine alignment for a resonator with more than three mirrors \cite{Jia:18PRA}. Our cavity has polarization eigenmodes with measured polarizations which are to a good approximation $R$ and $L$ with a measured frequency splitting of $\omega_R- \omega_L= 2\pi\times 0.9$ MHz, which is smaller than $\kappa$ and $\Gamma_e$ but unfortunately not much smaller. It is expected that this splitting can be reduced considerably in future experiments.

We choose the signal light during the control and the target pulse to be resonant with their respective cavity resonances. We choose the control signal light to be resonant with the atomic transition $\Delta_s= 0$. Hence, in the absence of target coupling light, the target signal light has a detuning of $\Delta_s/2\pi= -0.9$ MHz from the atomic transition. In the presence of target coupling light, the atomic ground state experiences a light shift of 0.18 MHz resulting in $\Delta_s/2\pi= -0.7$ MHz. We optimize the target coupling detuning to obtain a conditional $\pi$ phase shift, which turns out to be close to the two-photon resonance.

\section{Blockade Radius}

\label{app-blockade}

In principle, one could calculate the blockade radius in cavity Rydberg EIT near a F\"{o}rster resonance as
\begin{align}
R_\text{block}
= \left|\left(\frac{2C_3}{\hbar \Omega}\right)^2 \frac{\Gamma_e}{\gamma_F-2i\Delta_F} C\right|^{1/6}
\end{align}
with $C$, $\Omega$, and $\Gamma_e$ as in Eq.\ \eqref{C-eff}. Here, the signal and coupling light are assumed to be resonant with the atomic transitions $\Delta_s= \Delta_\text{co}= 0$ and the signal light is assumed to be resonant with the cavity $\Delta_c=0$.

The F\"{o}rster resonance comes about because the electric dipole-dipole interaction $V_{dd}$ couples the atom pair states $|\gamma,J,M\rangle= |48\,^2S_{1/2},50\,^2S_{1/2},J,M\rangle$ and $|\gamma',J', \linebreak[1] M'\rangle= |48\,^2P_{1/2},49\,^2P_{1/2},J',M'\rangle$, where $J,M$ and $J',M'$ are the quantum numbers of the total angular momentum and where the $z$ axis is chosen along the internuclear axis. This coupling is described by a matrix element $\langle \gamma,1,\pm 1|V_{dd}|\gamma',1,\pm 1\rangle= -C_3/R^3$, where $R$ is the internuclear distance, the $C_3$ coefficient is $C_3= 1.2\times10^6$ a.u., and one atomic unit is $6.460\times10^{-49}$ Jm$^3$. The rate coefficient $\gamma_F$ describes the off-diagonal decay of the density matrix according to $(\partial_t + \frac12 \gamma_F) \langle g,r'| \rho|\gamma'\rangle= 0$. For $R\to\infty$, the energies of states $|\gamma\rangle$ and $|\gamma'\rangle$ differ by the F\"{o}rster defect $\hbar\Delta_F$. In our experiment, this value vanishes by virtue of a static electric field which is applied using electrodes inside the vacuum chamber.

One can generalize this model to accommodate the random distribution of the angle included by the internuclear axes and the wave vector of the EIT signal light. One can combine this with the anisotropy of $V_{dd}$ characterized by $\langle \gamma,1,0|V_{dd}|\gamma',1,0\rangle= 2C_3/R^3$. This yields a small increase of $R_\text{block}$, which we neglect.

In practice, determining $\gamma_F$ is not easy. We temporarily assume that $\gamma_{rg}$ and $\gamma_F$ are both dominated by electric field noise. Hence, one naively expects $\gamma_F= |\frac{\alpha_{\gamma'}-\alpha_{r'}}{\alpha_r}|\gamma_{rg}$, where the static electric polarizabilities are $(\alpha_{r'},\alpha_{r},\alpha_{\gamma'})= (0.15,0.20,1.9)\times10^{12}$ a.u.\ for the states $(|r'\rangle,|r\rangle,|\gamma'\rangle)$. Here, one atomic unit is $1.649\times 10^{-41}$ J(m/V)$^2$, and we neglect the static ground-state polarizability of 319 a.u. Using $\gamma_{rg}$ from the fit to the blue data in Fig.\ \ref{fig-spectra}, we obtain $1/\gamma_F= 25(6)$ ns and $R_\text{block}= 7$ $\mu$m.

Note that, $\gamma_{rg}$ is not dominated by electric field noise because repeating the measurement of Fig.\ \ref{fig-spectra} at approximately zero electric field (so that the sensitivity to electric field fluctuations is greatly reduced) does not yield a big improvement regarding $\gamma_{rg}$. However, the value of $\gamma_F$ extracted from the above estimate is so large that it seems unlikely that other mechanisms produce a larger value of $\gamma_F$. Hence, we obtain the worst-case estimates $1/\gamma_F\geq 25(6)$ ns and $R_\text{block}\geq 7$ $\mu$m.

\section{Efficiency}

\label{app-efficiency}

\subsection{Average Efficiency}

Here, we study some mathematical properties of the average efficiency $\bar \eta$. To motivate the definition of $\bar \eta$, we first review a few properties of two widely used averages $\bar F$ and $F_\text{pro}$ for describing the fidelity between a real process $\mathcal E$ and an ideal process $\mathcal E_U$. In doing so, we assume that the ideal process is unitary, i.e.\ there exists a unitary operator $U$ such that $\mathcal E_U(\rho)= U\rho U^\dag$. On the one hand, we have the average fidelity \cite{Gilchrist:05}
\begin{align}
\label{F-bar}
\bar F
= \int d\psi \tr\big([\mathcal E_U(|\psi\rangle\langle\psi|)]^\dag \mathcal E(|\psi\rangle\langle\psi|)\big)
,\end{align}
where the integral is over the uniform (Haar) measure of all normalized state vectors. On the other hand, we have the process fidelity $F_\text{pro}$. For any orthonormal basis $B_i$ of operators, $F_\text{pro}$ can be expressed as the arithmetic mean
\begin{align}
\label{F-pro-sum}
F_\text{pro}(\mathcal E_U,\mathcal E)
= \frac1{d^2} \sum_{i=1}^{d^2} \tr\big([\mathcal E_U(B_i)]^\dag \mathcal E(B_i)\big)
,\end{align}
where $d$ is the dimension of the space of state vectors. This could serve as a definition of $F_\text{pro}$, although it is customary to start from a different definition \cite{Gilchrist:05}. Note that Eq.\ \eqref{F-pro-sum} has been stated e.g.\ in Eq.\ (10) in Ref.\ \cite{Gilchrist:05} but with the unnecessary additional restriction that each of the $B_i\sqrt d$ be unitary.

Both definitions, Eqs.\ \eqref{F-bar} and \eqref{F-pro-sum}, average the expression $\tr\{[\mathcal E_U(\rho)]^\dag \mathcal E(\rho)\}$ but while $\bar F$ averages over all normalized state vectors, $F_\text{pro}$ averages over an orthonormal basis of operators. Both types of averaging seem natural. Interestingly, the two definitions yield different results but are connected by the affine linear function \cite{Gilchrist:05}
\begin{align}
\label{F-bar-F-pro}
\bar F
= \frac{F_\text{pro}d+1}{d+1}
.\end{align}

In analogy to the definition of the average fidelity $\bar F$ in Eq.\ \eqref{F-bar}, we define the average efficiency
\begin{align}
\label{eta-bar-int}
\bar \eta
= \int d\psi \eta(|\psi\rangle\langle\psi|)
,\end{align}
where, again, the integral is over the uniform (Haar) measure of all normalized state vectors. We show below that, for any orthonormal basis $|v_i\rangle$ of state vectors, $\bar \eta$ can be expressed as the arithmetic mean
\begin{align}
\label{eta-bar-sum-2}
\bar \eta
= \frac1d \sum_{i=1}^d \eta(|v_i\rangle\langle v_i|)
= \frac1d \tr(\Theta)
.\end{align}
This expression is somewhat analogous to $F_\text{pro}$ in Eq.\ \eqref{F-pro-sum} but with an interesting difference, namely that the \emph{same} $\bar \eta$ appears in Eqs.\ \eqref{eta-bar-int} and \eqref{eta-bar-sum-2}, whereas $\bar F$ and $F_\text{pro}$ are connected by Eq.\ \eqref{F-bar-F-pro} in a less trivial manner. One of the main reasons why Eqs.\ \eqref{F-bar}, \eqref{F-pro-sum}, \eqref{eta-bar-int}, and \eqref{eta-bar-sum-2} are appealing is that the results are independent of the choice of basis.

As an alternative to $\bar\eta$, one could consider e.g.\ the harmonic mean of the eigenvalues of $\Theta$. This harmonic mean appears when applying the gate to produce a postselected Bell state with well-balanced intensities, as discussed in appendix \ref{app-improved-coincidence}. But in the long run, a more typical application consists in cascading many CNOT gates in a large network to perform a complicated quantum computation. Here, the relevant figure of merit in terms of efficiency is probably the average efficiency $\bar\eta$, not the harmonic mean of the eigenvalues of $\Theta$.

Now, we turn to the proof of Eq.\ \eqref{eta-bar-sum-2}. To this end, we use the following property of the uniform (Haar) measure, proven below,
\begin{align}
\label{int-psi-psi-psi}
\int d\psi |\psi\rangle\langle\psi|
= \frac1d \mathbbm 1
,\end{align}
where $\mathbbm 1$ is the identity matrix. Combining Eqs.\ \eqref{eta-bar-int} and \eqref{int-psi-psi-psi} with the linearity of $\eta$, we immediately obtain
\begin{align}
\label{eta-bar-identiy-matrix}
\bar \eta
= \eta\left(\frac1d \mathbbm 1\right)
.\end{align}
If the $|v_i\rangle$ form an arbitrary orthonormal basis of state vectors, then combining the linearity of $\eta$, the closure relation $\sum_{i=1}^d |v_i\rangle\langle v_i|= \mathbbm 1$, and Eq.\ \eqref{eta-bar-identiy-matrix} we immediately obtain Eq.\ \eqref{eta-bar-sum-2}.

It remains to be shown that Eq.\ \eqref{int-psi-psi-psi} holds. To see this, we first note that $A= \int d\psi |\psi\rangle\langle\psi|$ is a Hermitian operator. Hence, we choose an orthonormal basis of eigenvectors $|s_i\rangle$. In addition, we note that $UAU^\dag= A$ for all unitary $d\times d$ matrices $U$ because, for any fixed such $U$, we can substitute the integrand from $|\psi\rangle$ to $|\xi\rangle= U|\psi\rangle$ and use the defining property $\int d\psi \cdots = \int d\xi \cdots$ of the uniform (Haar) measure. As swapping eigenvectors $|s_i\rangle$ and $|s_j\rangle$ is a unitary operation, we find that all eigenvalues of $A$ are identical. Hence, there exists a complex number $c$ such that $A= c\mathbbm 1$. Considering $\tr(A)$ and using the fact that the uniform (Haar) measure is normalized, $\int d\psi = 1$, we obtain $c= 1/d$.

\subsection{Efficiency Tomography}

Efficiency tomography aims at determining $\Theta$ from a sufficiently large set of measurement outcomes. As $\eta$ is linear, it suffices to measure $\eta(A_j)$ for all elements $A_j$ of an arbitrary basis $\mathcal A$ of operators. To calculate the efficiency matrix $\Theta$ from the measured values $\eta(A_j)$, it is useful to introduce the dual basis with elements $A^j$ defined by $\tr(A_i^\dag A^j)= \delta_{i,j}$. The dual basis can be calculated according to
\begin{align}
A^j
= \tsum_{i=1}^{d^2} A_i h_{i,j}
,&&
h
= g^{-1}
,&&
g_{i,j}
= \tr(A_i^\dag A_j)
,\end{align}
where $g$ is a metric tensor. A more common notation for $h_{i,j}$ is $g^{i,j}$. For all operators $B$, this yields $B= \sum_{j=1}^{d^2} A^j \tr(A_j^\dag
\linebreak[1]
B)$. From this, one easily obtains
\begin{align}
\Theta
= \tsum_{j=1}^{d^2} A^j [\eta(A_j)]^*
.\end{align}

\section{GHZ-State Model}

\label{app-GHZ-model}

Here, we develop a simple model from which we derive Eqs.\ \eqref{p-H}--\eqref{C-N} which describe the observed values of $p_H$, $p_V$, $\mathcal P_N$, and $\mathcal C_N$. We consider an incoming state with exactly one control qubit and exactly $N-1$ target qubits. We assume that the state immediately after the gate, before applying the single-qubit unitary to all target qubits, is
\begin{multline}
\label{psi-GHZ}
|\psi\rangle
= \sqrt{1-\eta_N} |\psi_\text{loss} \rangle
+ \sum_{j\in\{H,V\}} e^{i\beta_c\delta_{j,V}} \frac{c_j}{\sqrt2} |j\rangle
\\
\otimes \left(\frac{t_{j,H}}{\sqrt2} |H\rangle + e^{i\beta_t} \frac{t_{j,V}}{\sqrt2} |V\rangle\right)^{\otimes (N-1)}
.\end{multline}
Here, the amplitudes $t_{H,V}= \mathcal R_b$, $t_{V,V}= \mathcal R$, and $t_{H,H}= t_{V,H}= 1$ model the target-qubit efficiency and the amplitudes $c_H= \sqrt{\eta_\text{sr}}$ and $c_V= \sqrt{\eta_f}$ model the control-qubit efficiency. In addition, $\eta_N$ denotes the probability that none of the incoming $N$ photons was lost. The normalized state $|\psi_\text{loss} \rangle$ represents terms in which at least one photon was lost.

Here, $\beta_c$ and $\beta_t$ denote single-qubit phases with shot-to-shot fluctuations, just as in Eq.\ \eqref{beta-c-beta-t-a} and \eqref{beta-c-beta-t-b}. For these random variables, we use the same assumptions as above and, again, we abbreviate $V_c= \overline{e^{i\beta_c}}$ and $V_t= \overline{e^{i\beta_t}}$.

We propagate all target qubits in the state $|\psi\rangle$ from Eq.\ \eqref{psi-GHZ} through the single-qubit unitary with $|D\rangle\mapsto |V\rangle$ and $|A\rangle\mapsto |H\rangle$. Next, we calculate the resulting density matrix and we average this density matrix over the random variables $\beta_c$ and $\beta_t$. We calculate the density matrix describing the subensemble postselected upon detection of exactly one control qubit and exactly $N-1$ target qubits. From here, we obtain Eqs.\ \eqref{p-H}--\eqref{C-N}.

\section{Quantum Process Tomography}

\label{app-tomography}

Quantum process tomography is described in detail in Ref.\ \cite{Nielsen:00}. Here, we point out a useful property of the below-defined matrix $\beta^{\mathcal D}$ which simplifies the mathematical procedure of quantum process tomography and, to our knowledge, has been overlooked in the literature so far.

The matrix representation $M^{\mathcal D}$ of the linear map $\mathcal E$ with respect to a basis of operators $\mathcal D:(D_1,\dots,D_{d^2})$ is defined by
\begin{align}
\mathcal E(D_j)
= \tsum_{i=1}^{d^2} D_i M^{\mathcal D}_{i,j}
.\end{align}
The matrix representation $M^{\mathcal D}$ is easily calculated from the results of quantum-state tomography for a basis of input density matrices. The map from $M^{\mathcal D}$ to $\mathcal E$ is an isomorphism, i.e.\ linear and bijective. According to Eq.\ \eqref{E-rho-chi}, the chi-matrix representation of $\mathcal E$ with respect to the same basis is defined by $\mathcal E(\rho)= \sum_{i,j=1}^{d^2} D_i \rho D_j^\dag \chi^{\mathcal D}_{i,j}$. The map from $\chi^{\mathcal D}$ to $\mathcal E$ is also an isomorphism. Hence, the map from $\chi^{\mathcal D}$ to $M^{\mathcal D}$ is an isomorphism as well. Thus, this map has a matrix representation $\beta^{\mathcal D}$ defined by \cite{Nielsen:00}
\begin{align}
M^{\mathcal D}_{i,j}
= \tsum_{k,\ell=1}^{d^2} \beta^{\mathcal D}_{i,j,k,\ell} \chi^{\mathcal D}_{k,\ell}
.\end{align}
If the basis $\mathcal D$ is orthonormal, one easily finds
\begin{align}
\label{beta-tr}
\beta^{\mathcal D}_{i,j,k,\ell}
= \tr(D_i^\dag D_k D_j D_\ell^\dag)
.\end{align}
Note the nontrivial ordering of the indices $j$ and $k$. Also note that $\beta^{\mathcal D}$ is independent of the quantum process $\mathcal E$.

As the map from $\chi^{\mathcal D}$ to $M^{\mathcal D}$ is an isomorphism, $\beta^{\mathcal D}$ is an invertible matrix. Note that $\beta^{\mathcal D}$ has $d^8$ matrix elements. For the two-qubit problem at hand, this gives $4^8= 65\,536$ matrix elements. Hence, inverting $\beta^{\mathcal D}$ can be cumbersome, even more so for larger quantum systems.

However, one can drastically simplify this matrix inversion by noting that if $\mathcal D$ is orthonormal, then $\beta^{\mathcal D}$ will be self-inverse, $\sum_{k,\ell=1}^{d^2}\beta^{\mathcal D}_{i,j,k,\ell}\beta^{\mathcal D}_{k,\ell,m,n}= \delta_{i,m} \delta_{j,n}$. To our knowledge, this useful relation has been overlooked in the literature so far. This relation is easy to prove in the basis of operators formed by the $|u_i\rangle\langle u_j|$, where the $|u_i\rangle$ form any orthonormal basis of state vectors. Considering a unitary change of basis of operators shows that this relation holds for all orthonormal bases of operators $\mathcal D$. This relation implies
\begin{align}
\label{chi-beta-M}
\chi^{\mathcal D}_{i,j}
= \tsum_{i,j,k,\ell=1}^{d^2} \beta^{\mathcal D}_{i,j,k,\ell} M^{\mathcal D}_{k,\ell}
,\end{align}
if $\mathcal D$ is orthonormal.

Performing a change of basis is typically much less cumbersome than inverting the large matrix $\beta^{\mathcal D}$. Hence, the best strategy for calculating the process matrix from the experimental data is typically to first calculate $M^{\mathcal D}$ with respect to some orthonormal basis $\mathcal D$, then calculate $\beta^{\mathcal D}$ using Eq.\ \eqref{beta-tr} and $\chi^{\mathcal D}$ using Eq.\ \eqref{chi-beta-M}, and, if needed, perform a final change of basis to obtain $\chi$ in any desired basis.

For completeness, we note that we choose the basis of input density matrices for quantum process tomography to consist of the 16 tensor product states which can be formed from the single-qubit state vectors $|H\rangle$, $|V\rangle$, $|D\rangle$, and $|R\rangle$.


\begin{thebibliography}{10}

\bibitem{Nielsen:00}
M.~A. Nielsen and I.~L. Chuang, {\em Quantum Computation and Quantum
  Information} (Cambridge University Press, Cambridge, 2000).

\bibitem{OBrien:07}
J.~L. O{\textquoteright}Brien, Optical quantum computing, {\em Science} {\bf
  318}, 1567--1570 (2007).

\bibitem{Kok:07}
P. Kok, W.~J. Munro, K. Nemoto, T.~C. Ralph, J.~P. Dowling, and G.~J. Milburn,
  Linear optical quantum computing with photonic qubits, {\em Rev. Mod. Phys.}
  {\bf 79}, 135--174 (2007).

\bibitem{Wehner:18}
S. Wehner, D. Elkouss, and R. Hanson, Quantum internet: {A} vision for the road
  ahead, {\em Science} {\bf 362}, aam9288 (2018).

\bibitem{Zhong:20}
H.-S. Zhong, H. Wang, Y.-H. Deng, M.-C. Chen, L.-C. Peng, Y.-H. Luo, J. Qin, D.
  Wu, X. Ding, Y. Hu, P. Hu, X.-Y. Yang, W.-J. Zhang, H. Li, Y. Li, X. Jiang,
  L. Gan, G. Yang, L. You, Z. Wang, L. Li, N.-L. Liu, C.-Y. Lu, and J.-W. Pan,
  Quantum computational advantage using photons, {\em Science} {\bf 370},
  1460--1463 (2020).

\bibitem{OBrien:03}
J.~L. O'Brien, G.~J. Pryde, A.~G. White, T.~C. Ralph, and D. Branning,
  Demonstration of an all-optical quantum controlled-{NOT} gate, {\em Nature}
  {\bf 426}, 264--267 (2003).

\bibitem{Kieling:10}
K. Kieling, J.~L. O'Brien, and J. Eisert, On photonic controlled phase gates,
  {\em New J. Phys.} {\bf 12}, 013003 (2010).

\bibitem{Knill:01}
E. Knill, R. Laflamme, and G.~J. Milburn, A scheme for efficient quantum
  computation with linear optics, {\em Nature} {\bf 409}, 46--52 (2001).

\bibitem{Li:21}
J.-P. Li, X. Gu, J. Qin, D. Wu, X. You, H. Wang, C. Schneider, S. H\"ofling,
  Y.-H. Huo, C.-Y. Lu, N.-L. Liu, L. Li, and J.-W. Pan, Heralded nondestructive
  quantum entangling gate with single-photon sources, {\em Phys. Rev. Lett.}
  {\bf 126}, 140501 (2021).

\bibitem{Hacker:16}
B. Hacker, S. Welte, G. Rempe, and S. Ritter, A photon-photon quantum gate
  based on a single atom in an optical resonator, {\em Nature} {\bf 536},
  193--196 (2016).

\bibitem{Tiarks:19}
D. Tiarks, S. Schmidt-Eberle, T. Stolz, G. Rempe, and S. D\"urr, A
  photon-photon quantum gate based on {Rydberg} interactions, {\em Nat. Phys.}
  {\bf 15}, 124--126 (2019).

\bibitem{Hao:15}
Y.~M. Hao, G.~W. Lin, K. Xia, X.~M. Lin, Y.~P. Niu, and S.~Q. Gong, Quantum
  controlled-phase-flip gate between a flying optical photon and a {Rydberg}
  atomic ensemble, {\em Sci. Rep.} {\bf 5}, 10005 (2015).

\bibitem{Das:16}
S. Das, A. Grankin, I. Iakoupov, E. Brion, J. Borregaard, R. Boddeda, I.
  Usmani, A. Ourjoumtsev, P. Grangier, and A.~S. S\o{}rensen, Photonic
  controlled-{PHASE} gates through {Rydberg} blockade in optical cavities, {\em
  Phys. Rev. A} {\bf 93}, 040303 (2016).

\bibitem{Pritchard:10}
J.~D. Pritchard, D. Maxwell, A. Gauguet, K.~J. Weatherill, M.~P.~A. Jones, and
  C.~S. Adams, Cooperative atom-light interaction in a blockaded {Rydberg}
  ensemble, {\em Phys. Rev. Lett.} {\bf 105}, 193603 (2010).

\bibitem{Parigi:12}
V. Parigi, E. Bimbard, J. Stanojevic, A.~J. Hilliard, F. Nogrette, R.
  Tualle-Brouri, A. Ourjoumtsev, and P. Grangier, Observation and measurement
  of interaction-induced dispersive optical nonlinearities in an ensemble of
  cold {Rydberg} atoms, {\em Phys. Rev. Lett.} {\bf 109}, 233602 (2012).

\bibitem{Firstenberg:13}
O. Firstenberg, T. Peyronel, Q.-Y. Liang, A.~V. Gorshkov, M.~D. Lukin, and V.
  Vuleti\'{c}, Attractive photons in a quantum nonlinear medium, {\em Nature}
  {\bf 502}, 71--75 (2013).

\bibitem{Jia:18}
N. Jia, N. Schine, A. Georgakopoulos, A. Ryou, L.~W. Clark, A. Sommer, and J.
  Simon, A strongly interacting polaritonic quantum dot, {\em Nat. Phys.} {\bf
  14}, 550--554 (2018).

\bibitem{Gyongyosi:21}
L. Gyongyosi and S. Imre, Scalable distributed gate-model quantum computers,
  {\em Sci. Rep.} {\bf 11}, 5172 (2021).

\bibitem{Calsamiglia:01}
J. Calsamiglia and N. L{\"u}tkenhaus, Maximum efficiency of a linear-optical
  {Bell}-state analyzer, {\em Appl. Phys. B} {\bf 72}, 67--71 (2001).

\bibitem{Guha:15}
S. Guha, H. Krovi, C.~A. Fuchs, Z. Dutton, J.~A. Slater, C. Simon, and W.
  Tittel, Rate-loss analysis of an efficient quantum repeater architecture,
  {\em Phys. Rev. A} {\bf 92}, 022357 (2015).

\bibitem{Reuer:22}
K. Reuer, J.-C. Besse, L. Wernli, P. Magnard, P. Kurpiers, G.~J. Norris, A.
  Wallraff, and C. Eichler, Realization of a universal quantum gate set for
  itinerant microwave photons, {\em Phys. Rev. X} {\bf 12}, 011008 (2022).

\bibitem{Raussendorf:01}
R. Raussendorf and H.~J. Briegel, A one-way quantum computer, {\em Phys. Rev.
  Lett.} {\bf 86}, 5188--5191 (2001).

\bibitem{Baur:14}
S. Baur, D. Tiarks, G. Rempe, and S. D\"urr, Single-photon switch based on
  {Rydberg} blockade, {\em Phys. Rev. Lett.} {\bf 112}, 073901 (2014).

\bibitem{Tiarks:14}
D. Tiarks, S. Baur, K. Schneider, S. D\"urr, and G. Rempe, Single-photon
  transistor using a {F\"orster} resonance, {\em Phys. Rev. Lett.} {\bf 113},
  053602 (2014).

\bibitem{Gorniaczyk:14}
H. Gorniaczyk, C. Tresp, J. Schmidt, H. Fedder, and S. Hofferberth,
  Single-photon transistor mediated by interstate {Rydberg} interactions, {\em
  Phys. Rev. Lett.} {\bf 113}, 053601 (2014).

\bibitem{Gorniaczyk:16}
H. Gorniaczyk, C. Tresp, P. Bienias, A. Paris-Mandoki, W. Li, I. Mirgorodskiy,
  H.~P. B\"{u}chler, I. Lesanovsky, and S. Hofferberth, Enhancement of
  {Rydberg}-mediated single-photon nonlinearities by electrically tuned
  {F\"{o}rster} resonances, {\em Nat. Commun.} {\bf 7}, 12480 (2016).

\bibitem{Gorshkov:07:cavity}
A.~V. Gorshkov, A. Andr\'e, M.~D. Lukin, and A.~S. S\o{}rensen, Photon storage
  in $\ensuremath{\Lambda}$-type optically dense atomic media. {I. Cavity}
  model, {\em Phys. Rev. A} {\bf 76}, 033804 (2007).

\bibitem{Hofmann:03}
H.~F. Hofmann, K. Kojima, S. Takeuchi, and K. Sasaki, Optimized phase switching
  using a single-atom nonlinearity, {\em J. Opt. B: Quantum and Semiclass.
  Opt.} {\bf 5}, 218--221 (2003).

\bibitem{Duan:04}
L.-M. Duan and H.~J. Kimble, Scalable photonic quantum computation through
  cavity-assisted interactions, {\em Phys. Rev. Lett.} {\bf 92}, 127902 (2004).

\bibitem{Gilchrist:05}
A. Gilchrist, N.~K. Langford, and M.~A. Nielsen, Distance measures to compare
  real and ideal quantum processes, {\em Phys. Rev. A} {\bf 71}, 062310 (2005).

\bibitem{OBrien:04}
J.~L. O'Brien, G.~J. Pryde, A. Gilchrist, D.~F.~V. James, N.~K. Langford, T.~C.
  Ralph, and A.~G. White, Quantum process tomography of a controlled-{NOT}
  gate, {\em Phys. Rev. Lett.} {\bf 93}, 080502 (2004).

\bibitem{Riebe:06}
M. Riebe, K. Kim, P. Schindler, T. Monz, P.~O. Schmidt, T.~K. K\"orber, W.
  H\"ansel, H. H\"affner, C.~F. Roos, and R. Blatt, Process tomography of ion
  trap quantum gates, {\em Phys. Rev. Lett.} {\bf 97}, 220407 (2006).

\bibitem{Tilma:02:SUN}
T. Tilma and E.~C.~G. Sudarshan, Generalized {Euler} angle parametrization for
  {SU(N)}, {\em J. Phys. A} {\bf 35}, 10467--10501 (2002).

\bibitem{Schmidt-Eberle:20}
S. Schmidt-Eberle, T. Stolz, G. Rempe, and S. D\"urr, Dark-time decay of the
  retrieval efficiency of light stored as a {Rydberg} excitation in a
  noninteracting ultracold gas, {\em Phys. Rev. A} {\bf 101}, 013421 (2020).

\bibitem{Sackett:00}
C.~A. Sackett, D. Kielpinski, B.~E. King, C. Langer, V. Meyer, C.~J. Myatt, M.
  Rowe, Q.~A. Turchette, W.~M. Itano, D.~J. Wineland, and C. Monroe,
  Experimental entanglement of four particles, {\em Nature} {\bf 404}, 256--259
  (2000).

\bibitem{Monz:11}
T. Monz, P. Schindler, J.~T. Barreiro, M. Chwalla, D. Nigg, W.~A. Coish, M.
  Harlander, W. H{\"a}nsel, M. Hennrich, and R. Blatt, 14-qubit entanglement:
  creation and coherence, {\em Phys. Rev. Lett.} {\bf 106}, 130506 (2011).

\bibitem{Wang:16}
X.-L. Wang, L.-K. Chen, W. Li, H.-L. Huang, C. Liu, C. Chen, Y.-H. Luo, Z.-E.
  Su, D. Wu, Z.-D. Li, H. Lu, Y. Hu, X. Jiang, C.-Z. Peng, L. Li, N.-L. Liu,
  Y.-A. Chen, C.-Y. Lu, and J.-W. Pan, Experimental ten-photon entanglement,
  {\em Phys. Rev. Lett.} {\bf 117}, 210502 (2016).

\bibitem{Li:15}
Y. Li, P.~C. Humphreys, G.~J. Mendoza, and S.~C. Benjamin, Resource costs for
  fault-tolerant linear optical quantum computing, {\em Phys. Rev. X} {\bf 5},
  041007 (2015).

\bibitem{Vaneecloo:22}
J. Vaneecloo, S. Garcia, and A. Ourjoumtsev, Intracavity {Rydberg} superatom
  for optical quantum engineering: {Coherent} control, single-shot detection,
  and optical $\ensuremath{\pi}$ phase shift, {\em Phys. Rev. X} {\bf 12},
  021034 (2022).

\bibitem{Saffman:10}
M. Saffman, T.~G. Walker, and K. M{\o}lmer, Quantum information with {Rydberg}
  atoms, {\em Rev. Mod. Phys.} {\bf 82}, 2313--2363 (2010).

\bibitem{Kuppens:00}
S.~J.~M. Kuppens, K.~L. Corwin, K.~W. Miller, T.~E. Chupp, and C.~E. Wieman,
  Loading an optical dipole trap, {\em Phys. Rev. A} {\bf 62}, 013406 (2000).

\bibitem{Urvoy:19}
A. Urvoy, Z. Vendeiro, J. Ramette, A. Adiyatullin, and V.
  Vuleti\ifmmode~\acute{c}\else \'{c}\fi{}, Direct laser cooling to
  {Bose-Einstein} condensation in a dipole trap, {\em Phys. Rev. Lett.} {\bf
  122}, 203202 (2019).

\bibitem{Tiarks:16}
D. Tiarks, S. Schmidt, G. Rempe, and S. D{\"u}rr, Optical $\pi$ phase shift
  created with a single-photon pulse, {\em Sci. Adv.} {\bf 2}, e1600036 (2016).

\bibitem{Gea-Banacloche:95}
J. Gea-Banacloche, Y.-q. Li, S.-z. Jin, and M. Xiao, Electromagnetically
  induced transparency in ladder-type inhomogeneously broadened media: {Theory}
  and experiment, {\em Phys. Rev. A} {\bf 51}, 576--584 (1995).

\bibitem{Jones:48}
R.~C. Jones, A new calculus for the treatment of optical systems. {VII}.
  properties of the {$N$}-matrices, {\em J. Opt. Soc. Am.} {\bf 38}, 671--685
  (1948).

\bibitem{Haase:06}
A. Haase, B. Hessmo, and J. Schmiedmayer, Detecting magnetically guided atoms
  with an optical cavity, {\em Opt. Lett.} {\bf 31}, 268--270 (2006).

\bibitem{Turnbaugh:21}
C. Turnbaugh, J.~J. Axelrod, S.~L. Campbell, J.~Y. Dioquino, P.~N. Petrov, J.
  Remis, O. Schwartz, Z. Yu, Y. Cheng, R.~M. Glaeser, and H. Mueller,
  High-power near-concentric {Fabry-P\'{e}rot} cavity for phase contrast
  electron microscopy, {\em Rev. Sci. Inst.} {\bf 92}, 053005 (2021).

\bibitem{Jia:18PRA}
N. Jia, N. Schine, A. Georgakopoulos, A. Ryou, A. Sommer, and J. Simon, Photons
  and polaritons in a broken-time-reversal nonplanar resonator, {\em Phys. Rev.
  A} {\bf 97}, 013802 (2018).

\end{thebibliography}
\end{document}